\documentclass[structabstract]{aa}
\usepackage{graphicx}
\usepackage{amssymb,amsmath}
\usepackage[colorlinks]{hyperref}
\usepackage{array}
\usepackage{aalongtable}
\usepackage{lscape}
\usepackage{natbib}
\usepackage{epstopdf}
\newcommand{\gr}{{$\gamma$-ray}}
\newcommand{\lsim}{{\lower.5ex\hbox{$\; \buildrel < \over \sim \;$}}}
\newcommand{\gsim}{{\lower.5ex\hbox{$\; \buildrel > \over \sim \;$}}}
\newcommand{\nupeak}{$\nu_{\rm peak}$}
\newcommand{\nufnupeak}{$\nu_{\rm peak}$f$_{\nu_{\rm peak}}$}

\begin{document}
   \title{2WHSP: A multi-frequency selected catalog of HE and VHE $\gamma$-ray blazars and blazar candidates}
   

   \author{
          Y.-L. Chang張淯翎  \inst{1,2,3}
          \and
          B. Arsioli  \inst{1,2,4} 
		  \and			  
		  P. Giommi   \inst{2,4}		  
		  \and 
          P.  Padovani \inst{5,6} 
          }
        
        \institute{Sapienza Universit\`a di Roma, ICRANet, Dipartimento di Fisica, Piazzale Aldo Moro 5, I-00185 Roma, Italy
         \and  Italian Space Agency, ASI, via del Politecnico snc, 00133 Roma, Italy 
         \and ICRANet, P.zza della Repubblica 10,65122, Pescara, Italy
         \and 
         ICRANet-Rio, CBPF, Rua Dr. Xavier Sigaud 150, 22290-180 Rio de Janeiro, Brazil
         \and
	    European Southern Observatory, Karl-Schwarzschild-Str. 2, D-85748 Garching bei M\"unchen, Germany
	    \and
	    Associated to INAF - Osservatorio Astronomico di Roma, via Frascati 33, I-00040 Monteporzio Catone, Italy \\
        \email{yuling.chang@asdc.asi.it,bruno.arsioli@asdc.asi.it,paolo.giommi@asdc.asi.it,ppadovan@eso.org}
        }
\titlerunning{2WHSP: A catalog of HE and VHE $\gamma$-ray blazars and blazar candidates}  

\abstract
{}
{High Synchrotron Peaked blazars (HSPs) dominate the \gr\ sky at energies larger than a few GeV; however, only a few hundred blazars of this type have been catalogued so far. 
In this paper we present the 2WHSP sample, the largest and most complete list of HSP blazars available to date, which is 
an expansion of the 1WHSP catalog of $\gamma$-ray source candidates off the Galactic plane.}
{We cross-matched a number of multi-wavelength surveys (in the radio, infrared and X-ray bands) and applied selection criteria based on the radio to IR and IR to X-ray spectral slopes. 
To ensure the selection of genuine HSPs we examined the SED of each candidate and estimated the peak frequency of its synchrotron emission (\nupeak)  
using the ASDC SED tool, including only sources with \nupeak~$> 10^{15}$~Hz (equivalent to \nupeak~$> 4$~eV).} 
{We have assembled the largest and most complete catalog of HSP blazars to date, which includes 1691 sources.
A number of population properties, such as infrared colours, synchrotron peak, redshift distributions, and $\gamma$-ray spectral properties, have been used to characterise 
the sample and maximize completeness.  We also derived the radio $\log$N-$\log$S distribution.
This catalog has already been used to provide seeds to discover new very high energy objects within {\it Fermi}-LAT data and to look for the counterparts of neutrino and
ultra high energy cosmic ray sources, showing its potential for the identification of promising high-energy \gr\, sources and multi-messenger targets.}
{}

 
\maketitle
%
\section{Introduction} 

Blazars are a class of radio-loud Active Galactic Nuclei (AGN) hosting a jet oriented at a small 
angle with respect to the line of sight \citep{Blandford1978, Antonucci1993, Urry1995}. 
The emission of these objects is non-thermal over most or the entire electromagnetic spectrum, from radio frequencies to hard $\gamma$-rays. 
The observed radiation shows extreme properties, mostly due to relativistic amplification effects. 
The observed Spectral Energy Distribution (SED) presents a general shape composed of two bumps, one typically located in the infrared (IR) and sometimes extending to the X-ray band and the other one in the hard X-ray to $\gamma$-rays. 
If the peak frequency of the synchrotron bump (\nupeak) in $\nu$ - $\nu$F$_{\nu}$ space is larger than $10^{15}$~Hz (corresponding to $\sim$ 4~eV), a source is usually called High Synchrotron Peaked (HSP) blazars\citep{Padovani1995,Abdo2010}. 
HSP blazars are also considered to be extreme sources since the Lorentz factor of the electrons radiating at the peak of the synchrotron bump $\gamma_{peak}$ are the highest within the blazar population, and likely of any other type of steady cosmic sources. 
Considering a simple SSC model where $\nu_{\rm peak}=3.2 \times 10^6  \gamma^{2}_{\rm peak} B \delta $ \citep[e.g.][]{Giommi2012a}, assuming $B=0.1$ Gauss and Doppler factor $\delta  =10$, HSPs characterized by $\nu_{\rm peak} $ ranging between $ 10^{15}$ and $\gsim  10^{18}$~Hz demand $\gamma_{\rm peak} \approx 10^4-10^6$. 

The typical two-bump SED of blazars and the high energies that characterize HSPs imply that these objects occupy a distinct position in the optical to X-ray spectral index ($\alpha_{\rm ox}$) versus the radio to optical spectral index ($\alpha_{\rm ro}$) colour-colour diagram \citep{Stocke1991}. 
Considering the distinct spectral properties of blazars over the whole electromagnetic spectrum, selection methods based on $\alpha_{\rm ox}$ and $\alpha_{\rm ro}$ have long been used to search for new blazars. 
For example, \citet{Schachter1993} discovered 10 new BL Lacs via a multi-frequency approach with radio, optical and X-ray data, and  
their BL Lac nature with optical spectra. 

HSP blazars play a crucial role in very high energy (VHE) astronomy. 
Observations have shown that HSPs are bright and  variable sources of high energy \gr\, photons (TeVCat)\footnote{http://tevcat.uchicago.edu} and that they 
are likely the dominant component of the extragalactic VHE background \citep{Padovani1993,Giommi2006,DiMauro2014,Giommi2015,Ajello2015}. 
In fact, most of the extragalactic objects detected so far above a few GeV are HSPs \citep[][see also TeVCat]{Giommi2009,Padovani2015a,Arsioli2015a,Fermi2fhl2015}. 
However, it is known that only a few hundred HSP blazars are above the sensitivity limits of currently available $\gamma$-ray surveys. 
For example, the 1WHSP catalog \citep[][hereafter Paper I]{Arsioli2015a}, which was the largest sample of HSP blazars when it was published, shows that 
out of the 992 objects in the sample, 299 have an associated \gr\, counterpart in the {\it Fermi} 1/2/3FGL catalogs. 
Nevertheless there is a considerable number of relatively bright HSPs which still lack a \gr\, counterpart. 
These are likely faint point-like sources at or below the {\it Fermi}-LAT, detectability threshold
 and were not found by the automated searches carried out so far. 
Indeed, \citet{Arsioli2016} have detected $\approx\,150$ new \gr\, blazars based on a specific search around bright WHSP sources, using over 7 years of {\it Fermi}-LAT Pass 8 data. 

In the most energetic part of the \gr\, band photons from high redshift sources are absorbed by the extragalactic background light (EBL) emitted by galaxies and quasars \citep{Dermer2011,Pfrommer2013,Bonnoli2015}. 
Therefore, the TeV flux can drop by a very large factor compared to GeV fluxes, making distant TeV sources much more difficult to detect. 
Paper I has shown that with the help of multi-wavelength analysis, HSP catalogs can provide many good candidates for VHE detection. 


The currently known HSP blazars are listed in catalogs such as the 5th {\it Roma-BZCAT} \citep[][hereafter 5BZCat]{Massaro2015}, the Sedentary Survey \citep{Giommi1999,Giommi2005,Piranomonte2007}, \citet{Kapanadze2013}, and Paper I.  
However, the number of known HSPs is still relatively small with less than $\approx 1000$ cataloged HSPs till now. 
Significantly enlarging the number of high energy blazars is important to better understand 
their role within the AGN phenomenon, and should shed light on the cosmological evolution of blazars, which is still a matter of debate.

The 5BZCat is the largest compilation of confirmed blazars, containing 3561 sources, around 500 of which are of the HSP type. 
It includes blazars discovered in surveys carried out in all parts of the electromagnetic spectrum  
and is also based on an extensive review of the literature and optical spectra. 
The Sedentary survey comprises 150 extremely high X-ray to radio flux ratio $(\log f_{\rm x}/f_{\rm r}\geq 3\times10^{-10}~{\rm erg}~{\rm cm}^{-2}~{\rm s}^{-1}~{\rm Jy}^{-1})$ HSP BL Lacs. 
The sample was obtained by cross-matching the RASS catalog of bright X-ray sources \citep{Voges1999} and the NVSS 1.4~GHz radio catalog \citep{Condon1998}. 
\citet{Kapanadze2013} built a catalog of 312 HSPs with flux ratio $(f_{\rm x}/f_{\rm r}\geq 10^{-11}~{\rm erg}~{\rm cm}^{-2}~{\rm s}^{-1}~{\rm Jy}^{-1})$ selected from various X-ray catalogs, the NVSS catalog of radio sources, and the first edition of the $Roma-BZCAT$ catalog \citep{Massaro2009}. 
The 1WHSP sample relied on a pre-selection based on Wide-field Infrared Survey Explorer (WISE) 
IR colours, SED slope criteria, and \nupeak~$>10^{15}$ Hz. It includes 992 known, newly-identified, and candidate high galactic latitude ($b>|20^\circ|$) HSPs. 

In a series of papers \cite{Massaro2011,DAbrusco2012,Massaro2012} showed that most blazars occupy a specific region of the IR colour-colour diagram, which they termed the blazar strip. 
In Paper I we extended the blazar strip in the WISE colour-colour diagram to include all the Sedentary Survey blazars and called it the {\it Sedentary WISE colour domain} (SWCD). 
The SWCD is wider than the WISE blazar strip since it contains some blazars whose host galaxy 
is very bright, such as MKN421 (2WHSP J110427.3+381230) and MKN 501 (2WHSP J165353.2+394536). 
We understood from previous work that many low-luminosity HSP blazars have the IR colours dominated by the thermal component of the host giant elliptical galaxy. 
Therefore, a selection scheme adopting IR colour restrictions may work effectively for selecting cases where the non-thermal jet component dominates the IR band but is less efficient for selecting galaxy dominated sources (since they are spread over a larger area in the IR colour-colour plot). 

In the present paper we extend the previous 1WHSP catalog to lower Galactic latitudes ($b>|10^\circ|$) building the larger and more complete 2WHSP catalog including over 1600 blazars expected to emit at VHE energies by means of multi-frequency data. 

\section{Building the largest sample of HSP blazars}
\label{building}
\subsection{Initial data selection by spatial cross-matching}
Blazars are known to emit electromagnetic radiation over a very wide spectral range, from radio 
to VHE photons. As discussed in Paper I, an effective way of building large blazar 
samples is to work with multi-frequency data, especially from all-sky surveys, and apply selection 
criteria based on spectral features that are known to be specific to blazar SEDs. 

We followed Paper I  and started by cross-matching the AllWISE whole sky infrared 
catalog \citep{Cutri2013} with three radio surveys \citep[NVSS, FIRST, and SUMSS:][]
{Condon1998,White1997,Manch2003}. To take into account the positional uncertainties associated with each target, we used matching radii of 0.3~arcmin for the NVSS and the SUMSS surveys and 0.1~arcmin for the FIRST catalog. 
Then we performed an internal match for all IR-radio sources to eliminate duplicate entries coming from the different radio catalogs.
Keeping only the best matches between radio and IR, we selected 2,137,505 objects. 

After that, we demanded all radio-IR matching sources to have a counterpart in one of the X-ray catalogs available to us \citep[RASS BSC and FSC, 1SWXRT and deep XRT GRB, 3XMM, XMM slew, Einstein IPC, IPC slew, WGACAT, Chandra, and BMW:][]{Voges1999,Voges2000,DElia2013,Puccetti2011,Rosen2015,Saxton2008,Harris1993,Elvis1992,White2000,Evans2010,Panzera2003}. Therefore we cross-matched the IR-radio subsample with each X-ray catalog individually, taking into account their positional errors. 
For instance, a radius of 0.1~arcmin was adopted for the cross-correlations (as in Paper I) unless the positional uncertainty of a source was reported to be larger than 0.1~arcmin, as e.g. in the case of many X-ray detections in the RASS survey.
In these cases, we used the 95\% uncertainty radius (or ellipse major axis)  of each source as maximum distance for the cross match.  
Some X-ray catalogs have a very wide range of positional uncertainties, thus we separated the data by positional errors and used different cross-matching radii for these X-ray catalogs. 
The radii used for cross-matching the IR-radio subsample with each X-ray catalog are reported in Table~\ref{Xraycross}. 
We also restricted the sample by Galactic latitude $|b|>10^\circ$ to avoid complications in the Galactic plane. 
We combined all the IR-radio-X-ray matching sources and applied an internal cross-check keeping only single IR sources within 0.1~arcmin radius; this procedure reduced the sample to 28,376 objects.

\begin{table} [h!]
\begin{center}
\begin{tabular}{ccc}
\hline\hline
Catalog&Error position&Cross-matched\\
& & radius\\
\hline
RASS&0-36 arcsec&0.6 arcmin\\
&$>$37 arcsec&0.8 arcmin\\
Swift 1SWXRT&0-5 arcsec&0.1arcmin \\
&$>$5 arcsec&0.2 arcmin\\
Swift deep XRT GRB&all data&0.2 arcmin\\
3XMM DR4&0-5 arcsec&0.1 arcmin\\
&$>$5 arcsec& 0.2 arcmin\\
XMM Slew DR6&all data&10 arcsec\\
Einstein IPC&all data&40 arcsec\\
IPC Slew&all data&1.2 arcmin\\
WGACAT2&all data&50 arcsec\\
Chandra&all data&0.1 arcmin\\
BMW&all data&0.15 arcmin\\
\hline\hline
\end{tabular}
\caption{The cross-matching radii of the X-ray catalogs.}
\label{Xraycross}
\end{center}
\end{table}

\subsection{Further selection based on broad-band spectral slopes}
\label{slope}
Here we take advantage of the fact that HSP blazars show radio to X-ray SEDs that distinguish them from any other type of extragalactic sources by imposing two constraints on the spectral slopes, namely: 

\begin{equation}
\begin{aligned}
~0.05&<\alpha_{1.4{\rm GHz}-3.4\mu{\rm m}}<0.45 \\
0.4&<\alpha_{4.6\mu{\rm m}-1{\rm keV}}< 1.1
\end{aligned}\label{eq:slope}
\end{equation}
where $\alpha_{\nu1- \nu2}=-\frac{\log(f_{\nu1}/f_{\nu2})}{\log(\nu_1/\nu_2)}$, 

that is the same conditions applied to the 1WHSP catalog, with the exception that here we do not apply the criterion $-1.0< \alpha_{3.4 \mu{\rm m}-12.0\mu{\rm m}} < 0.7$. 
This choice was necessary to prevent the loss of IR galaxy-dominated HSPs which could still be promising VHE candidates \citep[see][for details]{Massaro2011,Arsioli2015a}. 
The parameter ranges given above are derived from the shape of the SED of HSP blazars, which is assumed to be similar to those of three well-known bright HSPs, i.e. MKN 421, MKN 501, and PKS 2155$-$304 shown in Fig.~3 of Paper I, which also fit within the limiting slopes ($\alpha_{1.4{\rm GHz}-3.4\mu{\rm m}}$ and $\alpha_{4.6\mu {\rm m}-1{\rm keV}}$) used for the selection. 

By avoiding the application of the IR slope constraints used for the 1WHSP sample, we select more HSP candidates, reducing the incompleteness at low radio 
luminosities where the IR flux is often dominated by the host galaxy. 

\subsection{Deriving \nupeak\, and classifying the sources}
\label{DerivingPeak}
The final pre-selection led to a sample of 5,518 HSP-candidates, 922 of which are also 1WHSP sources. 
Note that this initial sample includes most of the HSP blazars that had to be added to the 1WHSP sample as additional previously known sources that were missed by the 
original selection procedure.
To refine and further improve the quality of the sample we used the ASDC SED builder 
tool\footnote{http://tools.asdc.asi.it/SED} to examine in detail all 5,518 candidates, accepting only 
those with SEDs that are consistent with that of genuine HSPs. 
Finally the synchrotron component of each object that passed our screening was fitted using a third degree polynomial function so as to estimate 
parameters such as \nupeak, and \nufnupeak, the energy flux at the synchrotron peak. 

The host galaxies of HSP blazars are typically giant ellipticals, and their optical and near IR flux sometimes dominate the SED in these bands. 
In order to only fit the synchrotron component of HSP blazars, it is crucial to distinguish the non-thermal nuclear radiation from the flux coming from 
the host galaxy. 
To do so we used the standard giant elliptical galaxy template of the ASDC SED builder tool to judge if the optical data points were due to the host galaxy or from non-thermal synchrotron radiation. 
If the source under examination had ultraviolet data (such as Swift-UVOT or GALEX measurements) it was straightforward to tell if there was non-thermal emission from the object. 

In addition, to avoid selecting objects with misaligned jets, which are expected to be radio-extended, the accepted spatial extension of the radio counterparts (as reported in the original catalogues) was limited to 1 arcmin. 
This procedure was carried out whenever possible, based on the 1.4~GHz radio image from NVSS, which includes the entire sky north of $\delta= -40^{\circ}$, similarly to what had been done for the 1WHSP catalog. 
We could also identify radio extended sources from their SED, since radio extended objects typically display a steep radio spectrum. All cases where we could find evidence of radio  (or X-ray, typically from clusters: see below) extension were eliminated from the sample. 
At the end of this process we only accepted objects with \nupeak~$> 10^{15}$~Hz \citep{Padovani1995}.

Clearly, most bright sources in the current list are also included in the 1WHSP catalog. Many of the new 
catalog entries are fainter sources or objects located at low Galactic latitudes ($10^{\circ} < | b |< 20^{\circ}$). 
In some cases the optical data were consistent with thermal emission from the host galaxy, and the few radio, IR, or X-ray measurements that could be related to 
non-thermal emission were very sparse.  Clearly, more multi-frequency data are 
needed for these sources. 

We still have a number of unclear cases due to the lack of good multi-frequency data.  We flagged them accordingly. 
In addition, since the positional accuracy in X-ray surveys is usually not as precise as that of optical or radio surveys, the position of the X-ray counterparts sometimes may be  20 to 40 arcsec away from the radio and optical counterparts, introducing more uncertainty.

Many of the 2WHSP candidates have been observed by SWIFT with multiple short exposures. 
To allow for a more accurate estimation of \nupeak\, and \nufnupeak\, we summed all the SWIFT XRT observations that were taken within a 3 week interval. 

\subsection{Avoiding X-ray contamination from cluster of galaxies}
\label{clusters}
Blazars are certainly not the only objects that emit X-rays. For instance, galaxy clusters also show X-ray emission that is, however, normally spatially extended with a spectrum that peaks at $\approx 1-3$~KeV resulting from the emission of giant clumps of hot and low density diffused gas \citep[$\approx10^8$ K and $\approx10^{-3}~\rm{atoms}/{\rm cm}^{3}$: ][]{Sarazin1988, Bohringer2010, PerezTorres2009}. 
Since blazars and radio galaxies are often located in clusters of galaxies, the X-rays from the hot gas, if not correctly identified, might cause the SED of the candidate 2WHSP source 
to look like that of a HSP object, introducing a source of contamination for our sample. 

To avoid this problem we carried out an extensive check of bibliographic references\footnote{For the cross-check with ADS references on each source we have used the Bibliographic Tool available on 
the ASDC website.} and catalogs of cluster of galaxies (e.g. ABELL, PGC, MCXC, ZW, WHL, etc), excluding cases where cluster emission could be responsible for the observed X-rays. 
In addition, we used Swift XRT imaging data (which are available for $\approx 60\%$ of our sample) to distinguish between X-ray emission from blazar jets, which is point-like in the XRT count maps, and that from the clusters, which is often extended. 
The same procedure was followed using XMM images, whenever these could be found in the public archive.
In addition, we cross-matched our sample with the positions of RASS extended sources and with those of the {\it Planck} catalog of Sunyaev-Zeldovich sources \citep{Planck2015}. 

Finally, we visually inspected optical images and the error circle maps built with the ASDC explorer 
tool\footnote{http://tools.asdc.asi.it} looking for targets that could be related to clusters of galaxies. 

\begin{figure}[h!]
\centering
\includegraphics[width=4cm,height=4cm]{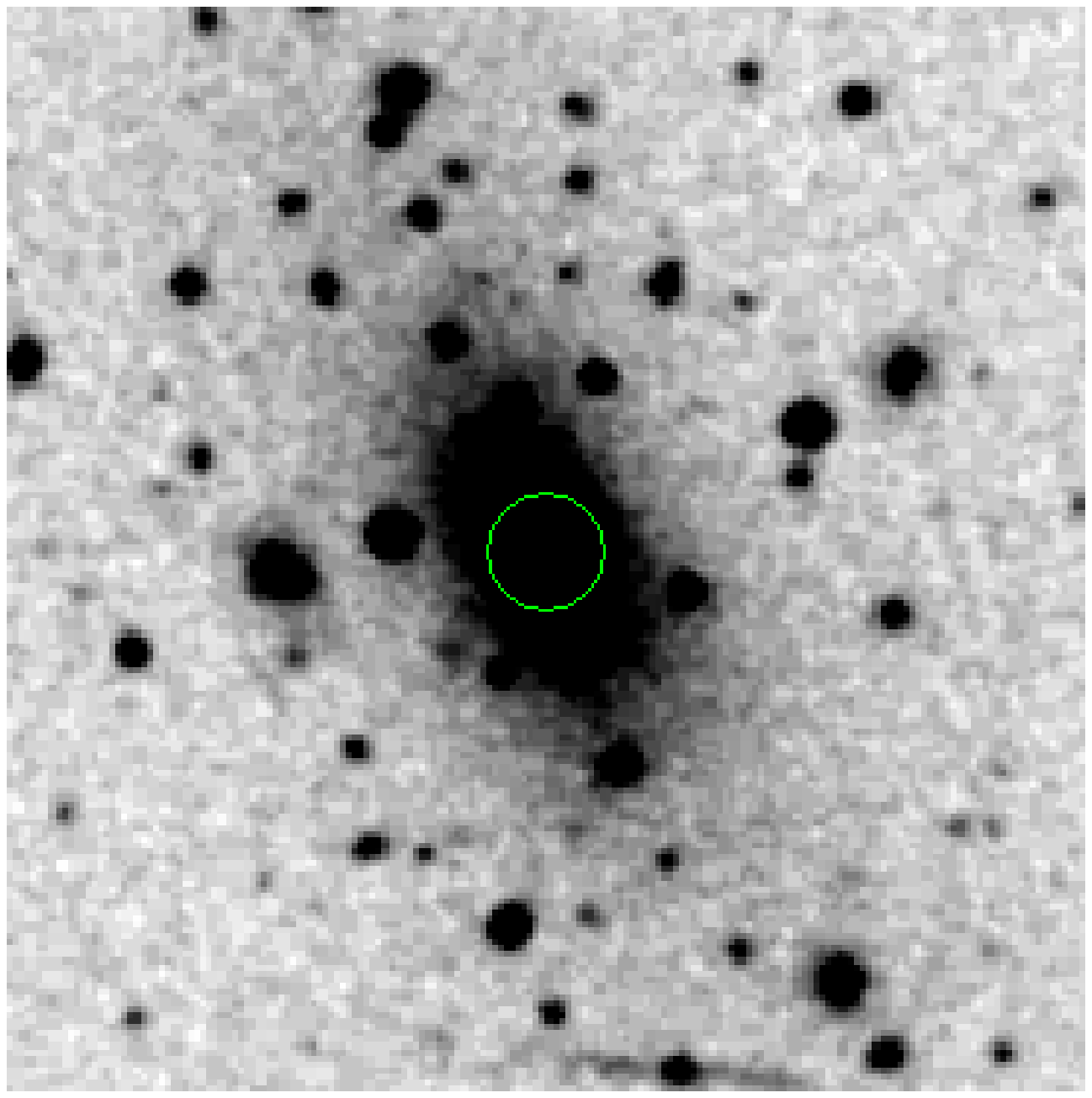} 
\includegraphics[width=4cm,height=4cm]{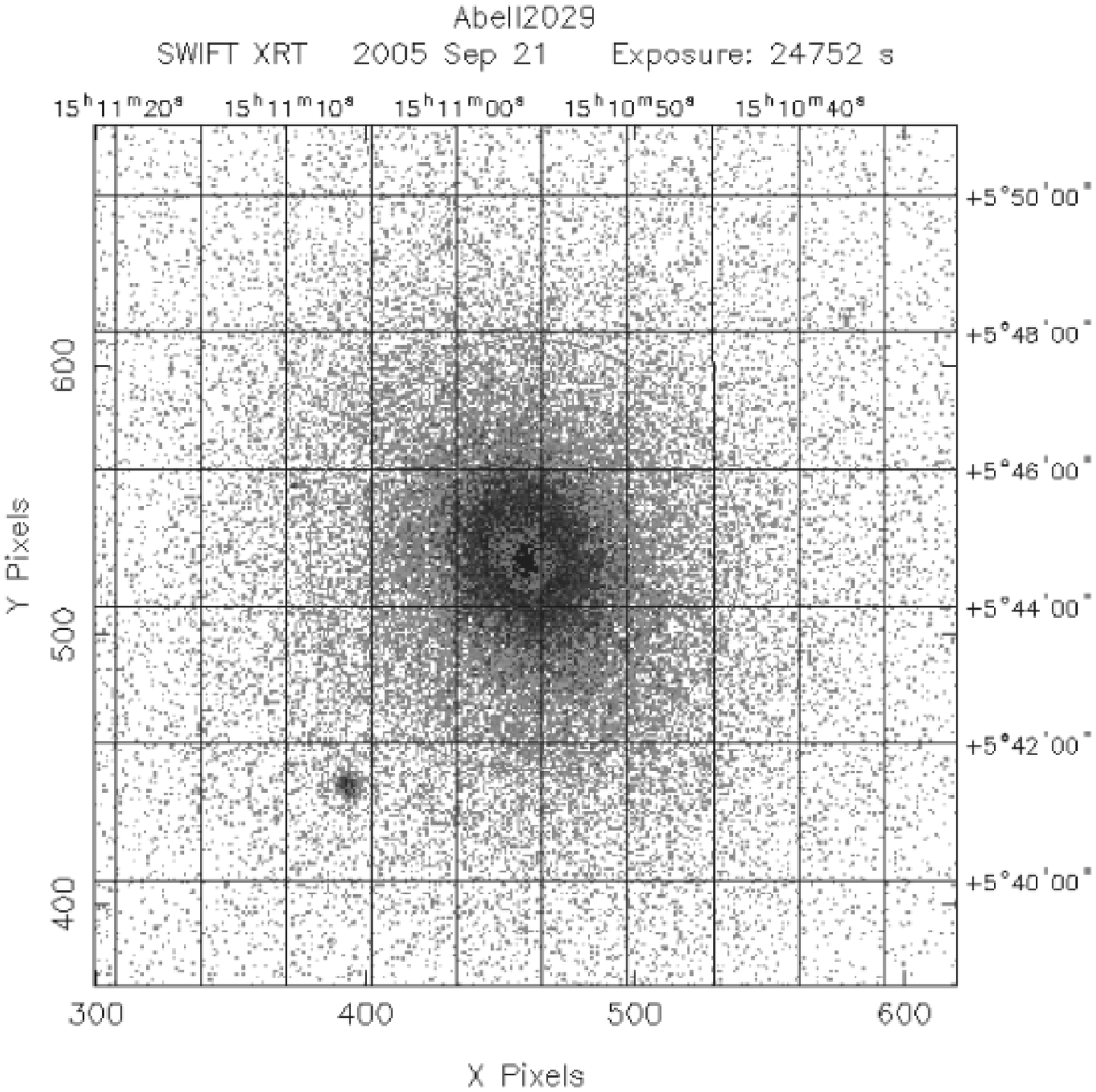} 
\caption[Abell2029]{Optical (left) and X-ray (right: XRT count map) iamges of 
WHL J151056.1+054441.}
\label{ext1}
\end{figure}

To illustrate how we removed objects that satisfy our multi-frequency selection criteria but where the X-ray flux is likely due to extended emission from a cluster of galaxies, 
we consider the example of WHL J151056.1+054441. 
This is a giant cluster of galaxies also cataloged as Abell2029. As the strong X-ray emission is clearly extended both in the Swift-XRT and XMM images  (see Fig. \ref{ext1}) 
 this source was removed from our HSP catalog. 

\begin{figure}[h!]
\centering
\includegraphics[width=4cm,height=4cm]{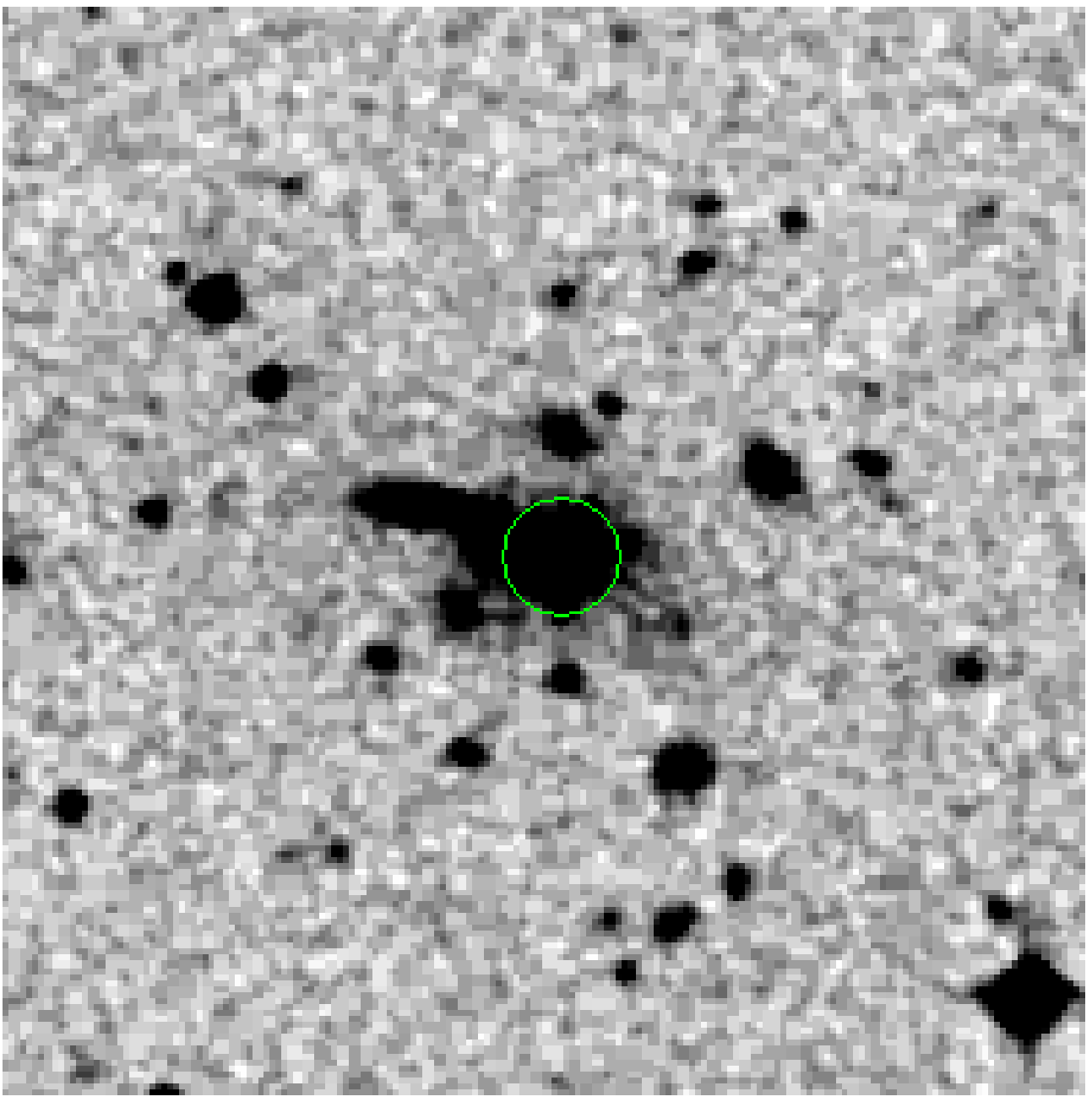} 
\includegraphics[width=4cm,height=4cm]{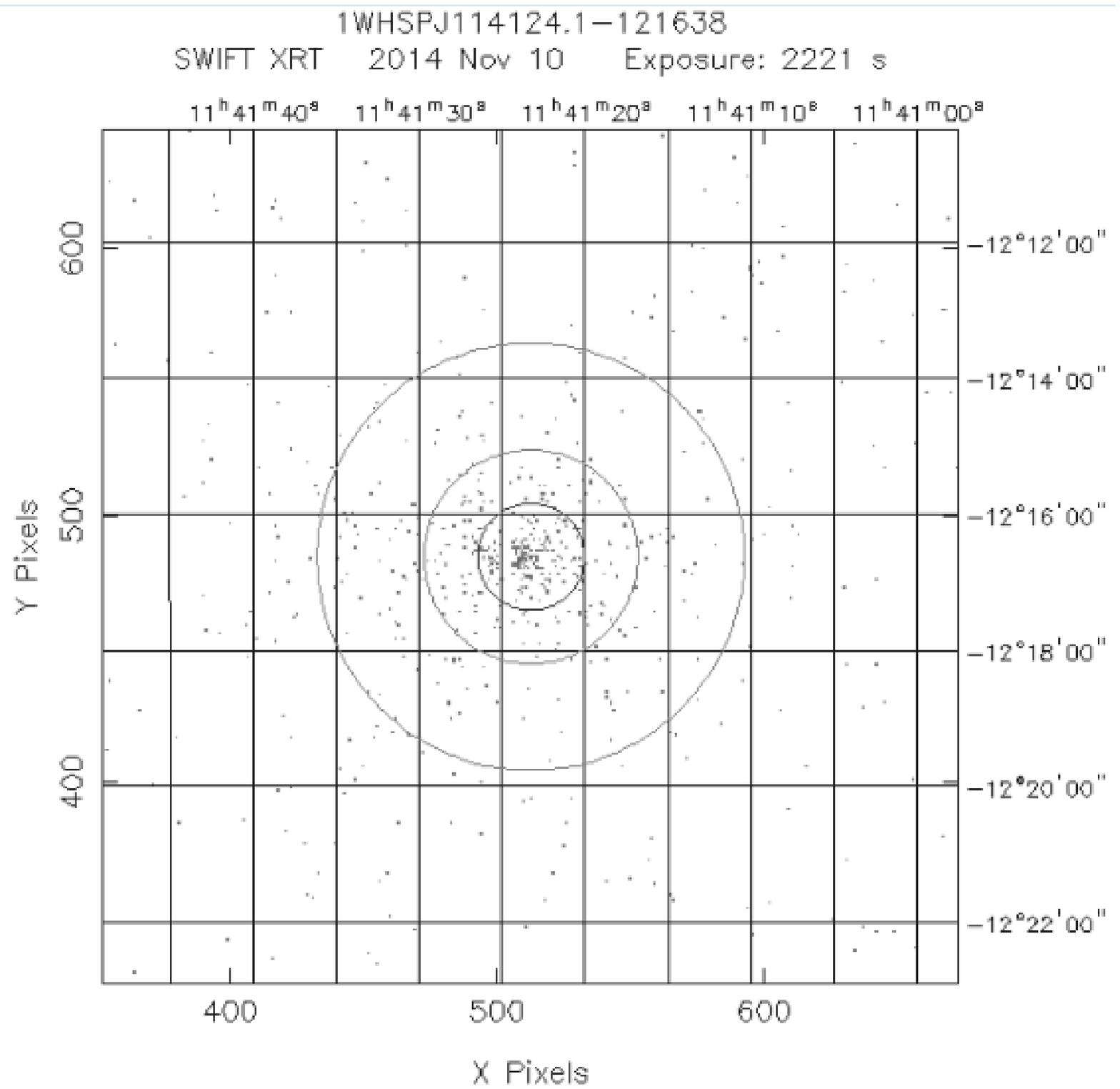} \\
\includegraphics[width=0.78\linewidth,angle=90]{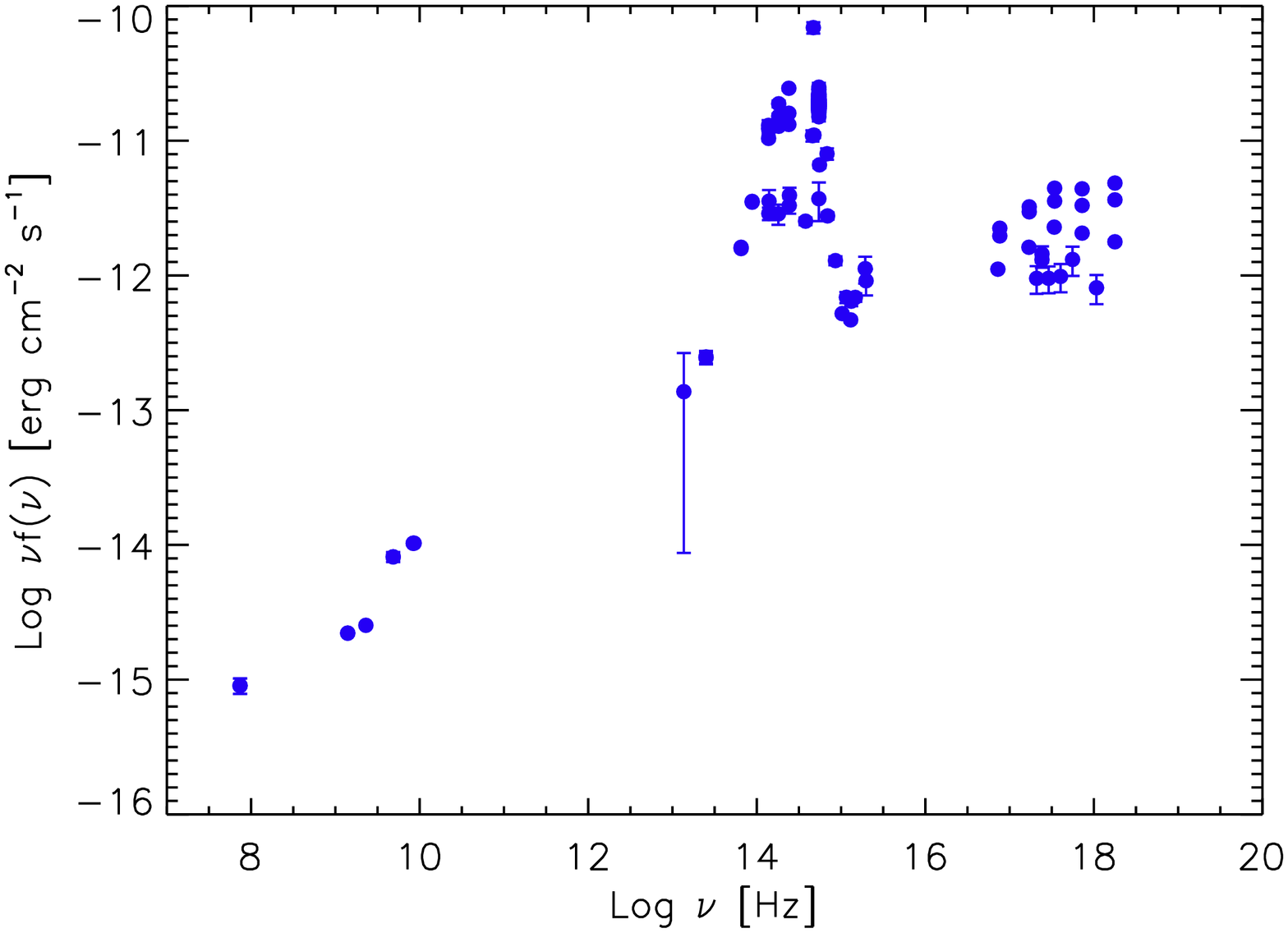}
\caption[LCRS B113851.7-115959]{Top: optical (left) and X-ray (right: XRT count map) images 
of LCRS B113851.7$-$115959. Bottom: the SED of LCRS B113851.7$-$115959.}
\label{ext2}
\end{figure}

Another example is shown in Fig. \ref{ext2}, where the candidate blazar is at the center of the cluster of galaxies LCRS B113851.7$-$115959. 
Although the X-ray emission is overall extended, the region around the sources shows clumps, and there are several X-ray detections; the non-thermal emission is very clear in the SED. Apparently, there is an AGN in the center that also emits in the UV. However, based on the available data we cannot know if the X-ray is mainly from the non-thermal jet or from the cluster 
and therefore we did not include this source in the catalog. 

\section{Improving the sample completeness}
\label{catalog}

The procedure described above led to the selection of 734 new HSPs in addition to those 
already included in the 1WHSP catalog, including previously known, newly discovered, and 
candidate blazars. For each source we adopted as best coordinates those taken from the 
WISE catalog. 

To evaluate the efficiency of our method of selecting VHE emission blazars, we cross-matched the sample of 1,647 objects with the Second Catalog of Hard {\it Fermi}-LAT Sources (2FHL) \citep{Fermi2fhl2015} and with TeVCat. 

Only 146 of the 360 sources in the 2FHL catalog (257 at $|b|>10^\circ$) are also in this preliminary sample. 
To verify if there are genuine HSPs in the 2FHL catalog that were missed by our selection, we closely examined the remaining 214 2FHL sources to see if they are 
cataloged as blazars. 
We found 31 high Galactic latitude blazars with \nupeak~$> 10^{15}$ Hz that could be added to the catalog. 
These sources were initially missed since they just did not match the optical-X-ray slope criteria (equation~\ref{eq:slope}) during the preliminary selection process. 
This selection inefficiency could be due to flux variability, lack of sufficiently high quality multi-frequency data, or simply to a non-optimal choice of parameter values 
in equation~\ref{eq:slope}.
Out of the 177 HSPs located at $|b|>10^\circ$ in the 2FHL catalog our selection method detected 146 objects, for an efficiency of $82.5\%$.

In addition, there are 14 HSP blazars in the 2FHL catalog that are located  at latitudes $|b|<10^\circ$, the area of sky that was not considered in our work 
to reduce complications connected to the Galactic plane. 
Since our aim is to provide the most complete list of HSPs we added to the 2WHSP catalog the 14 low latitude objects as well as all additional HSPs found in the 2FHL catalog, 
for a total of 45 sources. 
Only one good HSP blazar found among the 2FHL low Galactic latitude sources had no WISE data (2WHSP J135340.2$-$663958.0).
We used the radio position instead of the IR position in this case. 

We then checked catalogs of sources detected at TeV energies. 
Currently, the most complete list of objects detected in this band is TeVCat, which consists of 175 sources detected by Imaging Atmospheric/Air Cherenkov Telescope/Technique (IACT). 
At present there are three main IACT systems operating in the $\sim 50$~GeV to 50~TeV range: the High Energy Stereoscopic System (H.E.S.S.), MAGIC (Major Atmospheric Gamma Imaging Cherenkov Telescopes), and VERITAS (Very Energetic Radiation Imaging Telescope Array System). 
There are 38 TeVCat sources that are also in the 2WHSP catalog. 
We therefore checked the other high Galactic latitude sources to see if they were classified as HSP blazars, concluding that only one HSP source was missed. Note that previously we had already added 3 TeV sources to the 1WHSP catalog, since these were missed during its selection.
In total there are 39 HSPs at $|b|>10^\circ$ in TeVCat, 35 of which satisfy out selection criteria. 
Our selection efficiency in this case is $89.7 \%$. 

As in the case of the 2FHL catalog all missing sources have been lost because they just did not meet the slope criteria used in section~\ref{slope}. 
In all cases, however the spectral parameters turned out to be very close to the limits of the selection criteria, and \nupeak\, was $\approx 10^{15}$~Hz. 


The final 2WHSP catalog includes a total of 1691 sources, 288 of which are newly identified HSPs, 540 are previously known HSPs, 814 are 
HSP candidates, 45 are HSP blazars taken from the 2FHL catalog, and 4 from TeVcat. 
The complete list of 2WHSP sources is shown in Table~\ref{2whsptable}. 

We will further discuss the incompleteness due to the inefficiency in finding sources peaking at or just above  $10^{15}$ Hz in section~\ref{nupeak}. 


\section{Discussion}
\label{discussion}

\subsection{The \nupeak~distribution}
\label{nupeak}
\begin{figure}
\centering
\includegraphics[width=0.78\linewidth,angle =90]{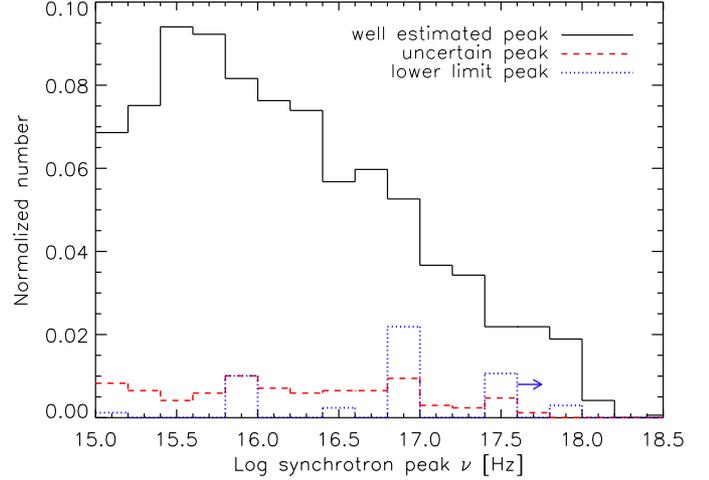}
\caption[The $\nu$ distribution]{The \nupeak\, distribution. 
The black solid line, blue dotted line, and red dashed line denote well estimated \nupeak, uncertain \nupeak, and lower limits on \nupeak, respectively. 
}
\label{nudist}
\end{figure}

The \nupeak~distribution of the 2WHSP sources is shown in Fig.~\ref{nudist}. 
The peak of the distribution is located at $\approx 10^{15.5}$~Hz and not at the threshold of \nupeak~ $= 10^{15}$~Hz used for the sample selection. 
This is very likely due to incompleteness of the sample near the \nupeak\, threshold, as 
our selection criteria were tuned to avoid too large an LSP contamination. 
The distribution is similar to that of the 1WHSP sample and of the subsample of HSP sources in the 5BZCat

When compared with other catalogs of extreme blazars, the peak value of the \nupeak~distribution of our sample is lower. For example the Sedentary and the \citet{Kapanadze2013}(hereafter K13) catalogs have peak values $\approx 10^{16.8}$ and $\approx 10^{16.7}$~Hz, respectively. 
This difference results from the criteria used and the different selected methods. 
The Sedentary and \citet{Kapanadze2013} catalogs, for example, were tuned to select sources 
with very large \nupeak\, values. 
Note that the  \nupeak\, of some sources is particularly high, with values $\gsim 10^{18}$~Hz. We discuss these extreme sources in the next section. 

 Sometimes, the severe variability of HSPs may result in displacements for \nupeak in different phases, such as MRK501 (See Fig.~\ref{extreme5}); not to mention that the intense variability will make the \nufnupeak vary 1-2 order or even worser. 
In these cases, we fit the \nupeak and \nufnupeak with the mean values in order to estimate the proper values for Synchrotron component averagely. 
By doing so, we avoid having extreme values for Synchrotron peak and reduce the effects of variability. 

\subsection{The highest \nupeak\, blazars}
\label{extremesec}

There are several sources in the 2WHSP sample with \nupeak\, around or above $10^{18}$Hz; these are usually called "extreme blazars". 
Values of \nupeak\, $\gsim 10^{18}$~Hz imply that the electrons responsible for the synchrotron radiation must be accelerated to extremely high energies
\citep[see the Introduction and e.g.][]{Rybicki1986,Costamante2001}. 

It is hard to estimate the positions of the synchrotron peak for such extreme sources, as the available data in the X-ray band is often limited to a few keV, where most of the sensitive existing detectors operate.
For about 60 sources we could not estimate well the frequency of the synchrotron peak since the soft X-ray data show a still rising spectrum in the SED,  
and no hard X-ray data exist to cover the peak of the emission. In these cases we could only estimate a lower limit to \nupeak.
For some strong X-ray variable sources with many X-ray observations we also could not obtain well-estimated \nupeak\, values with the third degree polynomial fitting in ASDC SED tool since the curvature in the X-ray spectrum (and with it \nupeak) changes with time. 
However, in all these cases the available multi-frequency data imply that the synchrotron peak is within the X-ray band; in these sources we estimated an average \nupeak\, value using a second-degree polynomial in the X-ray band.

Table~\ref{extremelist} gives the list of all the extreme sources with \nupeak~$\ge 10^{17.7}$~Hz; it includes many more such objects than any previous catalog. 
These extreme sources are particularly importance since they may be candidate VHE, neutrino or ultra high energy cosmic ray (UHECR) sources (section~\ref{vhecand} and \ref{neutrinocand}). 
Figure~\ref{extreme1} to \ref{extreme5} illustrate five examples of SEDs of representative objects from Table~\ref{extremelist}. 

\begin{itemize}
\item {\bf 2WHSP J023248.5+201717 (1ES0229+200)}. 
This is an extreme source with VHE data available 
\citep[the ebl-deabsorbed VHE data shown as black filled circles are from ][]{Finke2015}. 
The synchrotron peak is at $\sim 10^{18}-10^{19}$ Hz and the peak flux is one of the highest among the 2WHSP sources.  
In the VHE band, once one corrects the VHE fluxes for EBL absorption, the inverse Compton peak 
will be at energies $>1$~TeV. 

\begin{figure}
\centering
\includegraphics[width=1\linewidth]{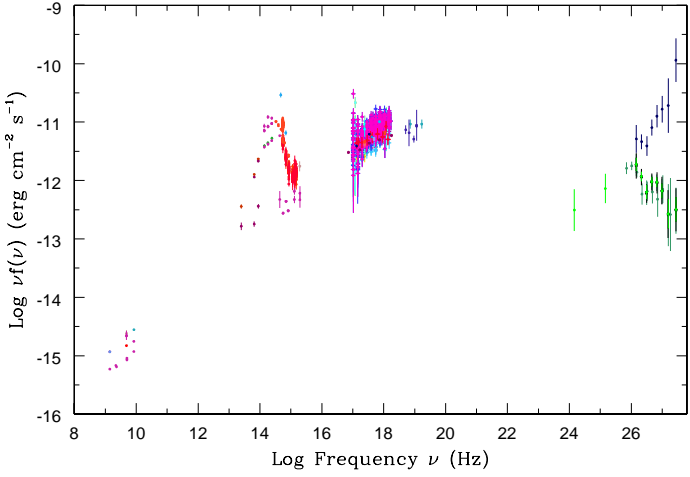}
\caption{The SED of the extreme object 2WHSP J023248.5+201717. The dark blue points are ebl- deabsorbed data from \citet{Finke2015}. See text for details.}
\label{extreme1}
\end{figure}

\item {\bf 2WHSP J035257.4$-$683117}. 
This is a known blazar with $\log$~\nupeak~$\approx 18.1$. 
It has hard X-ray and $\gamma$-ray detections but no TeV detection yet. 
This source might be a good target for next generation TeV telescopes. 
This source is not in 5BZCat yet.

\begin{figure}
\centering
\includegraphics[width=1\linewidth]{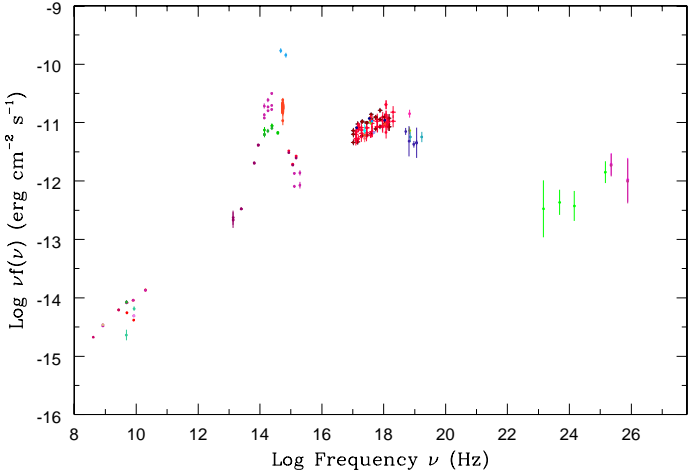}
\caption{The SEDs of the extreme object 2WHSP J035257.4$-$683117. See text for details.}
\label{extreme2}
\end{figure} 

\item {\bf 2WHSP J215305.2$-$004229 (5BZBJ2153$-$0042)}. 
This source has a very hard X-ray spectrum and the SED in the X-ray band keeps increasing 
up to the highest energies, implying a \nupeak\, larger that $10^{18}$~Hz. 
The X-ray emission is not likely to be related to a cluster of galaxy as it is compact. 
It has $\gamma$-ray data and may be a good TeV candidate source. 

\begin{figure}
\centering
\includegraphics[width=1\linewidth]{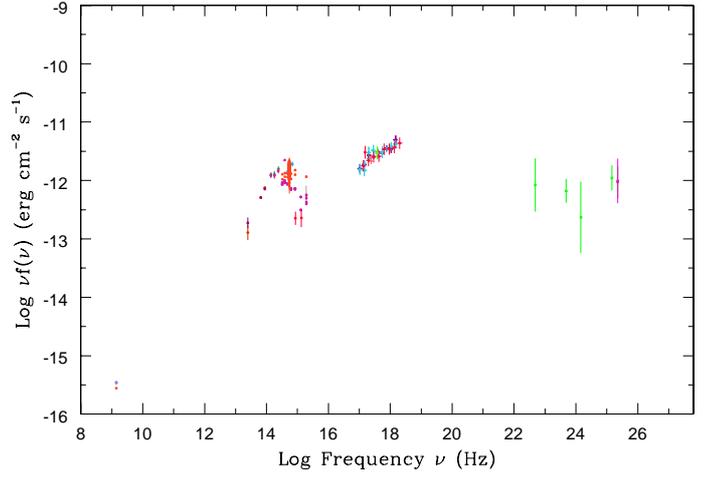}
\caption{The SED of the extreme object 2WHSP J215305.2$-$004229. See text for details.}
\label{extreme3}
\end{figure}

\item {\bf 2WHSP J143342.7$-$730437}. 
This is another example of a very hard X-ray SED. 
It has UV data but did not have any $\gamma$-ray data yet; however, this source is in the list of 
new $\gamma$-ray detections in \cite{Arsioli2016}. 

\begin{figure}
\centering
\includegraphics[width=1\linewidth]{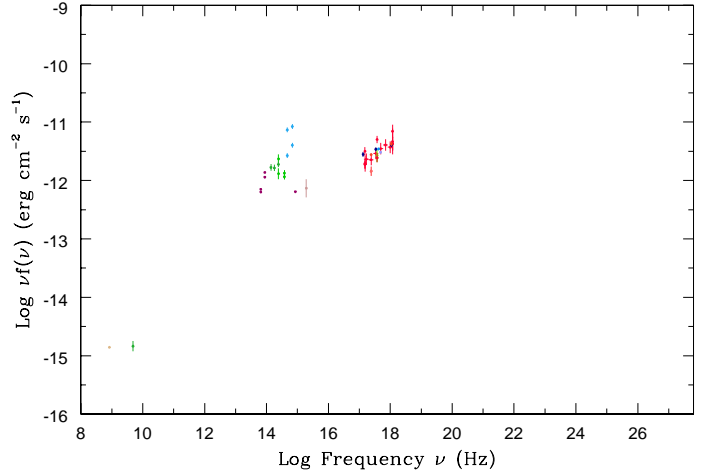}
\caption{The SEDs of the extreme object 2WHSP J143342.7$-$730437. See text for details.}
\label{extreme4}
\end{figure}

\item {\bf 2WHSP J165352.2+394536}. 
This is the well-known HSP MRK501. On average 
$\log \nu_{\rm peak} \sim 17.9$~Hz; however, during an X-ray flare, as shown by the BeppoSAX data (yellow points in the SED, \citet{Giommi2002}), \nupeak\, reached 
$>10^{18}$~Hz. 
Note that in \citet{Pian1998}, they discussed the BeppoSAX observation of MRK501 in 
April, 1997 and showed that the \nupeak of that shift at least two orders of magnitude w.r.t. previous observations of that. 
The scenario is seen for the first time at that time.

\begin{figure}
\centering
\includegraphics[width=1\linewidth]{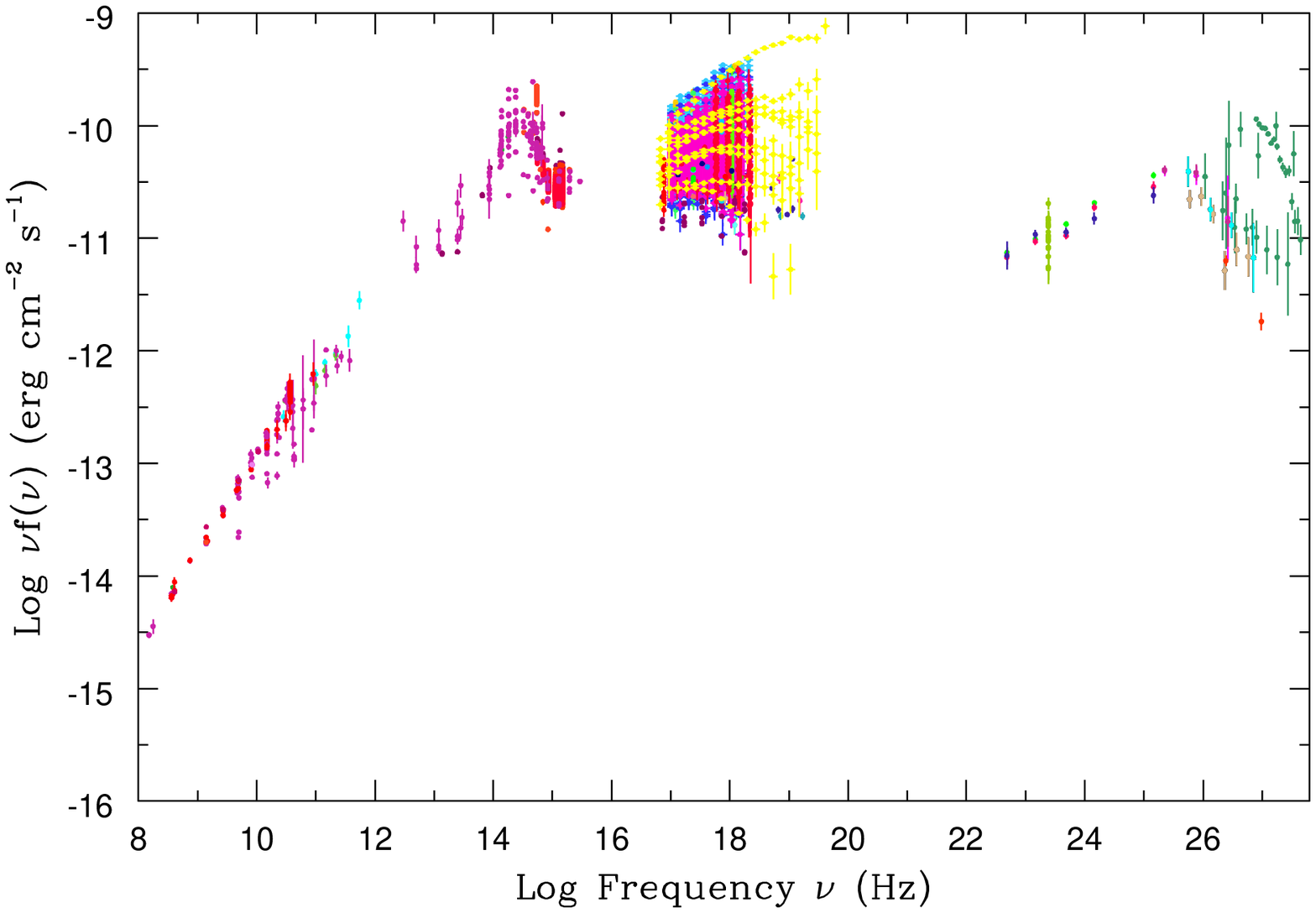}
\caption{The SEDs of the extreme object 2WHSP J165352.2+394536. See text for details.}
\label{extreme5}
\end{figure}

\end{itemize}

\begin{center}
\begin{table*}
\begin{tabular}{>{\centering\arraybackslash}p{4cm}>{\raggedleft\arraybackslash}p{1.6cm}>{\raggedleft\arraybackslash}p{1.8cm}>{\centering\arraybackslash}p{8cm}}
\hline\hline
Source&$\log$\nupeak&$\log$\nufnupeak&note\\
\hline
2WHSPJ003322.3$-$203907&17.9&-11.9&new HSP\\	
2WHSPJ004013.7+405003&$>$17.5&$>$-11.5&5BZU, lower limit\\
2WHSPJ012308.5+342048&18.0&-10.8&5BZB, TeV source\\
2WHSPJ013803.7$-$215530&$>$17.5&$>$-12.0&blazar candidate lower limit\\
2WHSPJ015657.9$-$530159&18.0&-11.1&5BZB \\
2WHSPJ020412.9$-$333339&17.9&-11.7&5BZB, new $\gamma$-ray identification\\ 
2WHSPJ023248.5+201717&18.5&-11.0&5BZG, TeV source*\\ 
2WHSPJ032056.2+042447&17.9&-11.7&blazar candidate, new $\gamma$-ray identification\\
2WHSPJ032356.5$-$010833&$>$17.5&$>$-11.9&5BZB, TeV source, lower limit\\
2WHSPJ034923.1$-$115926&17.9&-11.0&5BZB, TeV source\\
2WHSPJ035257.4$-$683117&18.1&-11.0& previously known BL Lac*\\
2WHSPJ050419.5$-$095631&17.9&-11.6&new HSP, new $\gamma$-ray identification\\
2WHSPJ050709.2$-$385948&$>$17.5&$>$-12.2&blazar candidate, lower limit\\
2WHSPJ050756.0+673723&17.9&-10.7&5BZB, TeV source\\
2WHSPJ055040.5$-$321615&18.1&-10.7&5BZG, TeV source\\
2WHSPJ055716.7$-$061706&17.9&-11.5&blazar candidate, new $\gamma$-ray identification\\
2WHSPJ064710.0$-$513547&17.9&-11.2&blazar candidate\\
2WHSPJ071029.9+590820&18.1&-10.7&5BZB, TeV source\\
2WHSPJ073326.7+515354&17.9&-11.3&blazar candidate\\
2WHSPJ081917.5$-$075626&18.0&-11.5&5BZB, TeV source\\
2WHSPJ083251.4+330011&18.0&-12.0&5BZB, new $\gamma$-ray identification\\
2WHSPJ084452.2+280409&17.9&-12.3&new HSP\\
2WHSPJ092057.4$-$225720&$>$17.5&$>$-11.6&new HSP, lower limit\\
2WHSPJ094620.2+010450&17.9&-11.8&5BZB, TeV source\\
2WHSPJ095849.0+013218&17.9&-12.3&new HSP, new $\gamma$-ray identification\\ 
2WHSPJ102212.6+512359&18.2&-11.7&5BZG, new $\gamma$-ray identification\\
2WHSPJ104651.4$-$253544&$>$18.0&$>$-11.5&5BZB\\
2WHSPJ105606.6+025213&17.9&-11.5&5BZG\\
2WHSPJ110357.1+261117&17,9&-12.2&new HSP\\
2WHSPJ110651.7+650603&17.9&-12.7&blazar candidate\\
2WHSPJ110804.9+164820&17.9&-12.7&new HSP\\
2WHSPJ112313.2$-$090424&17.9&-12.4&blazar candidate\\
2WHSPJ113209.1$-$473853&$>$17.5&$>$-11.6&blazar candidate, lower limit\\
2WHSPJ113630.1+673704&18.1&-11.1&5BZB, TeV source\\
2WHSPJ121323.0$-$261806&17.9&-11.2&5BZB\\
2WHSPJ122044.5+690525&$>$17.5&$>$-12.0&blazar candidate, lower limit\\
2WHSPJ122208.6+030718&$>$17.5&$>$-11.8&new HSP, lower limit\\
2WHSPJ122514.2+721447&$>$17.5&$>$-11.8&lower limit, 5BZB\\
2WHSPJ125341.2$-$393159&17.9&-11.3&5BZG, new $\gamma$-ray identification\\
2WHSPJ125708.2+264924&$>$17.5&$>$-12.3&new HSP, lower limit \\
2WHSPJ132239.1+494336&$>$17.5&$>$-12.1&new HSP, lower limit\\
2WHSPJ132541.8$-$022809&17.9&-12.0&5BZB, new $\gamma$-ray identification\\
2WHSPJ140027.0$-$293936&$>$17.5&$>$-12.1&blazar candidate, lower limit\\
2WHSPJ140121.1+520928&$>$17.5&$>$-12.0&5BZB, lower limit\\
2WHSPJ142832.5+424020&18.1&-10.7&5BZB, TeV source\\
2WHSPJ143342.7$-$730437&$>$17.5&$>$-11.5&blazar candidate, lower limit, new $\gamma$-ray identification*\\
2WHSPJ151041.0+333503&$>$17.5&$>$-11.5&5BZG, lower limit, new $\gamma$-ray identification\\
2WHSPJ151618.7$-$152344&18.0&-11.7&5BZB, new $\gamma$-ray identification\\
2WHSPJ153646.6+013759&$>$18.0&$>$-11.7&5BZB\\
2WHSPJ160519.0+542058&17.9&-12.0&5BZB, new $\gamma$-ray identification\\ 
2WHSPJ161004.0+671026&$>$17.5&$>$-11.8&5BZB, lower limit, new $\gamma$-ray identification\\
2WHSPJ161414.0+544251&17.9&-12.6&blazar candidate\\
2WHSPJ161632.8+375603&18.0&-12.1&5BZG\\2WHSPJ161632.8+375603&18.0&-12.1&5BZG\\
2WHSPJ162330.4+085724&$>$17.5&$>$-12.1& new HSP, lower limit, new $\gamma$-ray identification\\
2WHSPJ165352.2+394536&17.9&-10.2&Variability, flaring, 5BZB, TeV source*\\
2WHSPJ171902.2+552433&17.9&-12.5&known blazar\\
2WHSPJ194333.7$-$053352&$>$17.5&$>$-11.8&blazar candidate\\
2WHSPJ194356.2+211821&18.1&-11.0&new HSP, TeV source\\
2WHSPJ205528.2$-$002116&$>$18.0&$>$-10.9&5BZB, TeV source, lower limit\\
2WHSPJ214410.0$-$195559&17.9&-12.4&blazar candidate\\
2WHSPJ215305.2$-$004229&$>$18.0&$>$-11.4&5BZB, lower limit*\\
2WHSPJ223248.7$-$202226&17.9&-11.7&blazar candidate\\
2WHSPJ225147.5$-$320611&$>$18.0&$>$-11.3&5BZU, lower limit, new $\gamma$-ray identification\\
\hline\hline \\
\end{tabular}
\caption{The extreme synchrotron peaked sources. 
The sources marked with * are discussed in the text and shown in Figure \ref{extreme1} to \ref{extreme5}.}
\label{extremelist}
\end{table*}
\end{center}

\subsection{The redshift distribution}
Some 2WHSP sources lack redshift as their optical spectra are completely featureless. 
As in Paper I, we estimated lower limit redshifts for these sources. 
Assuming that in the optical band the host galaxy is swamped by the non-thermal emissions and leaves no imprint on the optical spectrum when the observed non-thermal flux is at least ten times larger than the host galaxy flux, we used the distance modulus (for details, see eq. 5 in Paper I) to calculate the lower limits redshifts. 
For the others, we obtained the redshifts from the references listed in Table~\ref{2whsptable}. 

\begin{figure}
\centering
\includegraphics[width=0.78\linewidth,angle =90]{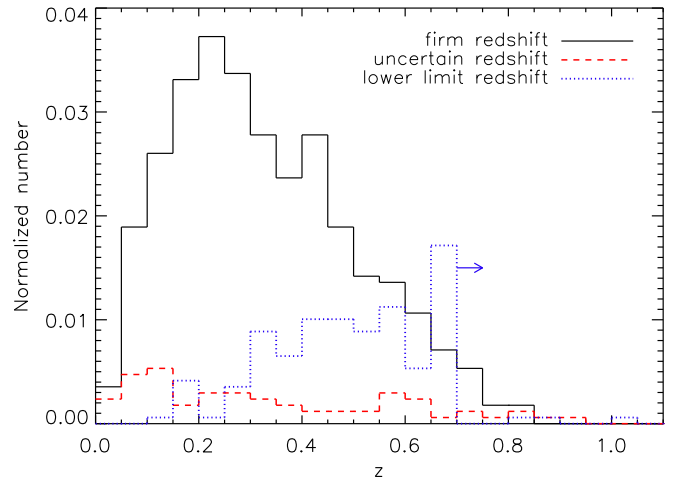}
\caption[The redshift distribution]{The redshift distribution of 2WHSP sources. 
The black solid line represents the sources with firm redshifts, the red dashed line the sources with uncertain redshift, and the blue dotted line the lower limits. 
}
\label{zdist}
\end{figure}

Fig.~\ref{zdist} shows the redshift distribution, which peaks just above 0.2. 
For all 2WHSP sources, $\langle z_{all} \rangle=0.371 \pm 0.005$; for firm redshift 2WHSP sources, $\langle z \rangle=0.331 \pm 0.008$. 
Clearly, sources without firm redshift are on average farther away than sources with firm redshift.
High redshift sources in flux limited samples tend to have featureless optical spectra as the host galaxy contribution is overwhelmed by the synchrotron emission. 
\citet{Giommi2012a} have predicted that the redshift distribution of BL Lacs without redshift in radio flux limited surveys will peak around $z_{predict} \approx 1.2$. 
The results again suggest that all source with only lower limit redshift or without redshift could be much further away than objects with measured redshift. 

Considering only sources with firm z values, the redshift distribution of 2WHSP sources is similar but
not identical to other HSP catalogs/subsamples. 
The average redshift of the 1WHSP catalog is $\langle z_{1whsp}\rangle=0.306$, that of the 
subsample of HSPs (\nupeak~$>10^{15}$~Hz) in 5BZCat is $\langle z_{bzcat}\rangle=0.294$, that 
of the Sedentary sources is $\langle z_{s}\rangle=0.320$, and that of the K13 catalog is 
$\langle z_{k}\rangle=0.289$. 
For instance in K13, the redshifts range is $0.031<z_{k}<0.702$, while in this paper we selected a number of sources with relatively high redshift ($z >0.7$) that are not in previous catalogs.

\subsection{The radio $\log$N-$\log$S of HSP blazars}
\label{lognlogssec}

The estimation of the statistical properties, such as the $\log$N-$\log$S  of a population of sources, requires the availability of flux limited and complete samples.  
As we demand that all 2WHSP sources have a radio, IR and X-ray counterpart, we must take into account the incompleteness resulting from 
the fact that the only existing all sky X-ray survey is not sufficiently deep to ensure the detection of all radio and IR faint HSP blazars. 

For the purpose of estimating the $\log$N-$\log$S we then considered the subsample of 2WHSP sources that are included in 
the RASS X-ray survey, which covers the entire sky albeit with sensitivity that strongly depends on ecliptic latitude (see Sec. 4.3 of Paper I for more details).

For each source in the 2WHSP-RASS subsample we therefore calculated a contribution $``{\rm n}_{\rm i}"$ to the total density, as given by ${\rm n}_{\rm i}=1/{\rm A}_{\rm i}~{\rm deg}^{-2}$, 
where the parameter ${\rm A}_{\rm i}$ is the sky area covered by RASS with sensitivity sufficient to detect the source in consideration.
We then sum the contribution of all sources in a given flux bin ${\rm N}_{\rm bin}=\sum {\rm n}_{\rm i}$ and obtain the $\log$N-$\log$S.
We use this approach to estimate the $\log$N-$\log$S of HSP blazars with respect to the radio flux density and the flux at the peak of the synchrotron component \nufnupeak. 

\begin{figure}
\centering
\includegraphics[width=0.78\linewidth,angle =90]{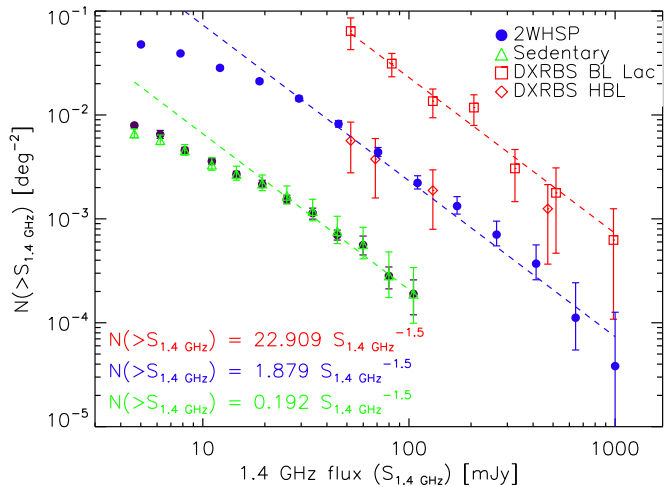}
\caption[The integral radio 1.4 GHz LogN-LogS]{The integral radio $\log$N-$\log$S at 1.4 GHz. 
The blue filled circles denote the 2WHSP catalog, the green open triangles indicate the Sedentary 
one, the red open squares represent DXRBS BL Lacs of all types, while the red diamonds 
are the subsample of HBLs in the DXRBS \citep[those in the HBL box: see ][for details]{Padovani2007}. The dashed lines have a 
slope of $-1.5$.}
\label{rrlognlogs}
\end{figure}

The integral radio $\log$N-$\log$S for the 2WHSP sample is shown in Fig.~\ref{rrlognlogs} where we also plot the $\log$N-$\log$S for the Sedentary HBL 
\citep{Giommi1999,Giommi2005,Piranomonte2007} for comparison. 
The dotted lines correspond to a fixed slope of -1.5, the expected value for a complete sample of a non-evolving population in a Euclidean Universe. 
Since the radio surveys that we use have different sensitivities in the northern and southern sky, we considered only sources with $\delta > -40^{\circ}$ and radio flux density $\ge 5$ mJy.

It is clear from Fig.~\ref{rrlognlogs} that the surface density of the 2WHSP sample is approximately a 
factor of ten larger than that of the Sedentary survey, which is expected since the latter includes 
more extreme sources (its \nupeak\, distribution peaks at $\log \nu_{\rm peak} \sim$ 16.8, 
as compared to $\log \nu_{\rm peak} \sim$ 15.5 for the 2WHSP sample). Apart from the different 
normalizations the $\log$N-$\log$S of the two samples show similar trends deviating from 
the Euclidean slope at radio flux densities lower than $\approx 20$~mJy. The 2WHSP flattening, however, appears 
to be stronger than the one of the Sedentary survey, which suggests the onset of some degree of incompleteness
at lower radio flux densities, on top of the evolutionary effects discussed by \cite{Giommi1999}. 
The 2WHSP maximum surface density corresponds
to a total of $\sim 1,900$ HSP blazars over the whole sky. Given that this number refers only to sources
with 1.4 GHz flux density $\ge 5$ mJy, and because of the incompleteness discussed above, 
this has to be considered a robust lower limit. 

Fig.~\ref{rrlognlogs} shows also the 5 GHz\footnote{Given that BL Lacs typically have flat
radio spectra we did not convert the 5 GHz counts to 1.4 GHz.} number counts for the Deep 
X-ray Radio Blazar Survey (DXRBS) BL Lacs (red squares) and HBL only (red diamonds) from 
\cite{Padovani2007}. The latter are in very good agreement with the 2WHSP number counts in the
region of overlap, which shows that our selection criteria are robust. Moreover, one can see a clear
trend going from the Sedentary survey to the 2WHSP sample and to the whole BL Lac population,
with an increase in number of a factor $\approx 10$ at every step. Given the unbiased nature of radio selection
with respect to \nupeak\, this is a direct consequence of BL Lac demographics, with HBL making up only $\sim 10\%$ 
of the total \citep[see also, e.g.][]{Padovani2007}. 


\subsection{The IR Colour-Colour plot}
\begin{figure}
\centering
\includegraphics[width=0.78\linewidth,angle =90]{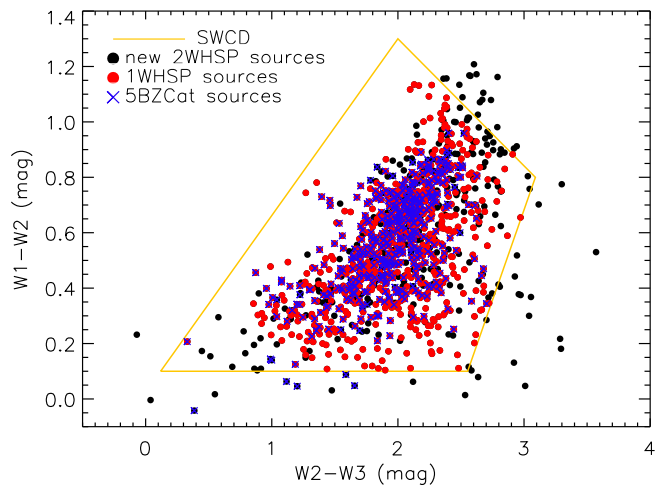}
\caption[The colour-colour diagram]{The IR colour-colour diagram. The black ones are the sources we selected in 2WHSP but not in 1WHSP, the red ones are the selected in 1WHSP, the blue crosses are the sources also in 5BZCat. The yellow line marks the SWCD region.}
\label{color}
\end{figure}

Figure~\ref{color} shows the WISE IR colour-colour diagram of 2WHSP sources, with signal to noise ratio (snr) in the W3 channel larger than 2, and the sources in the first WHSP sample and HSP blazars in the 5BZCat list. 
As expected, all of the 1WHSP sources are within the SWCD region as this was one of the criteria of the selection.  
By dropping the IR slope criterion ($-1.0< \alpha_{3.4 \mu {\rm m}-12.0\mu {\rm m}} < 0.7$) the 2WHSP sample includes more HSPs 
than the 1WHSP that are located in the bottom-left region within the SWCD.  

There are also 49 sources outside the SWCD region (see Table~\ref{swcdout}), six of them also in 5BZCat. 
The sources at the bottom are dominated by the host galaxy in the optical and near IR bands (Class 1). 
The right part of Fig.~\ref{color} is occupied by sources with problematic W3 photometry and sources 
whose W3 magnitude has relatively small snr values (typically $< 4$: Class 2). 
The sources located in the upper-right region have W1 fluxes similar or slightly lower than the W2 fluxes (Class 3). 
The class 3 sources may be IR variable sources or could be blazars at the border between ISP and HSP objects or 
might simply have poor W1 or W2 photometry. 
All 49 sources were checked individually and all of them are good HSP candidates. 
Thus, we suggest that the SWCD region needs to be extended to include all galaxy dominated HSPs. 

\begin{table*} 
\begin{center}
\begin{tabular}{crrrr}
\hline\hline
Source&W1 mag&W2 mag&W3 mag&W3 SNR\\
\hline
Class 1: Host galaxy dominated& & & & \\
2WHSPJ180408.8+004221&12.197&12.109&10.521&12.3\\
2WHSPJ085730.1+062726&13.349&13.303&12.101&2.7\\
2WHSPJ031250.2+361519&12.137&12.089&10.432&12.5\\
2WHSPJ090802.2$-$095936&11.586&11.523&10.406&15.0\\
2WHSPJ160740.0+254113&11.401&11.443&11.057&8.7\\
2WHSPJ013626.5+302011&15.961&15.914&12.905&2.2\\
2WHSPJ023109.1$-$575505&10.546&10.515&9.039&38.3\\
2WHSPJ085958.6+294423&14.881&14.802&12.166&3.0\\
2WHSPJ094537.0$-$301332&14.864&14.850&12.317&3.1\\
2WHSPJ120850.5+452951&14.811&14.749&12.629&2.6\\
2WHSPJ130711.8+115316&12.430&12.413&11.864&4.2\\
2WHSPJ195020.9+604750&12.675&12.679&12.640&3.2\\
2WHSPJ101514.2$-$113803&12.694&12.462&12.532&2.4\\
\hline
Class 2: Mainly problematic W3& & & & \\
2WHSP J000552.9$-$284502&15.501&15.370&12.450&2.3\\
2WHSP J004501.4+051215&14.170&13.795&10.822&3.1\\
2WHSP J082030.7$-$031412&15.118&14.676&11.749&3.0\\
2WHSP J082355.6+394747&15.509&15.141&12.086&3.3\\
2WHSP J100520.4+240503&15.324&14.946&11.988&3.0\\
2WHSP J113405.8+483903&15.108&15.005&12.422&2.4\\
2WHSP J122304.9+453444&15.246&14.841&12.033&4.0\\
2WHSP J122944.5+164004&15.265&14.965&11.925&3.2\\
2WHSP J124430.7+351002&15.782&15.080&11.964&3.6\\
2WHSP J140125.3+031629&16.550&15.775&12.476&3.1\\
2WHSP J144446.0+474256&15.840&15.427&12.630&3.2\\
2WHSP J162939.4+701448&16.763&16.233&12.662&3.2\\
2WHSP J175955.2+150109&15.907&15.690&12.404&2.8\\
2WHSP J195134.7$-$154929&14.702&14.417&11.715&3.4\\
2WHSP J212233.7+192527&15.201&14.683&11.739&4.9\\
2WHSP J215355.8$-$295443&15.914&15.733&12.440&2.2\\
\hline
class 3: W2 similar to or brighter than W1& & & & \\
2WHSP J002258.9$-$244022&15.056&13.894&11.104&8.4\\
2WHSP J022941.1$-$412050&14.773&13.747&11.113&9.0\\
2WHSP J024743.3$-$481545&15.164&14.059&11.382&10.5\\
2WHSP J025057.1$-$122612&15.081&14.001&11.332&8.3\\
2WHSP J054504.3+065809&14.953&14.049&11.129&5.9\\
2WHSP J071625.6+750700&15.768&14.634&12.147&5.3\\
2WHSP J093938.5$-$031502&15.328&14.445&11.535&5.0\\
2WHSP J095518.4$-$294611&14.321&13.149&10.526&14.2\\
2WHSP J120136.0$-$060733&15.247&14.168&11.430&3.5\\
2WHSP J135043.7$-$310926&14.359&13.202&10.817&12.5\\
2WHSP J172746.3$-$754618&14.039&12.883&10.522&15.3\\
2WHSP J180158.9+610938&15.332&14.164&11.598&9.9\\
2WHSP J185550.8+805223&16.492&15.580&12.638&3.4\\
2WHSP J202803.5+720513&15.440&14.459&11.671&13.1\\
2WHSP J204734.9+793759&16.494&15.328&12.753&2.9\\
2WHSP J213533.7+314919&14.312&13.223&10.711&12.0\\
2WHSP J233207.6$-$025245&15.108&14.037&11.387&5.0\\
2WHSP J233630.4$-$635634&14.599&13.391&10.788&12.0\\
\hline\hline
\end{tabular}
\caption{Sources outside the SWCD region}
\label{swcdout}
\end{center}
\end{table*}

\subsection{Candidates for GeV and VHE $\gamma$-ray observations}
\label{vhecand}
Since HSPs are the dominant population in the extragalactic VHE sky the 2WHSP catalog 
provides good candidates for the search of sources in {\it Fermi} catalogues and in the
VHE band. The Figure of Merit \citep[FOM, defined in][as the ratio between the synchrotron peak flux 
\nufnupeak~of a given source and that of the faintest blazar in the 1WHSP sample that has already 
been detected in the TeV band]{Arsioli2015a} was introduced to provide a simple quantitative measure of 
potential detectability of HSPs by TeV instruments.
The FOM parameter is reported for all 2WHSP sources and gives an objective way to assess the likelihood that a given HSP may be detectable as a TeV source. 
As discussed in Paper I, relatively high FOM sources (FOM $> 0.1$) are good targets for observation with the upcoming Cherenkov Telescope Array (CTA). Another upcoming instrument, the Large High Altitude Air Shower Observatory (LHAASO), is currently designed to survey the whole northern sky for $\gamma$-ray sources above 300~GeV, with unprecedented sensitivity.
Therefore, high FOM  2WHSP sources may also provide seed-positions for searches of $\gamma$-ray signature embedded in LHAASO data \citep{Cao2010}. 

\begin{figure}
\centering
\includegraphics[width=0.78\linewidth,angle=90]{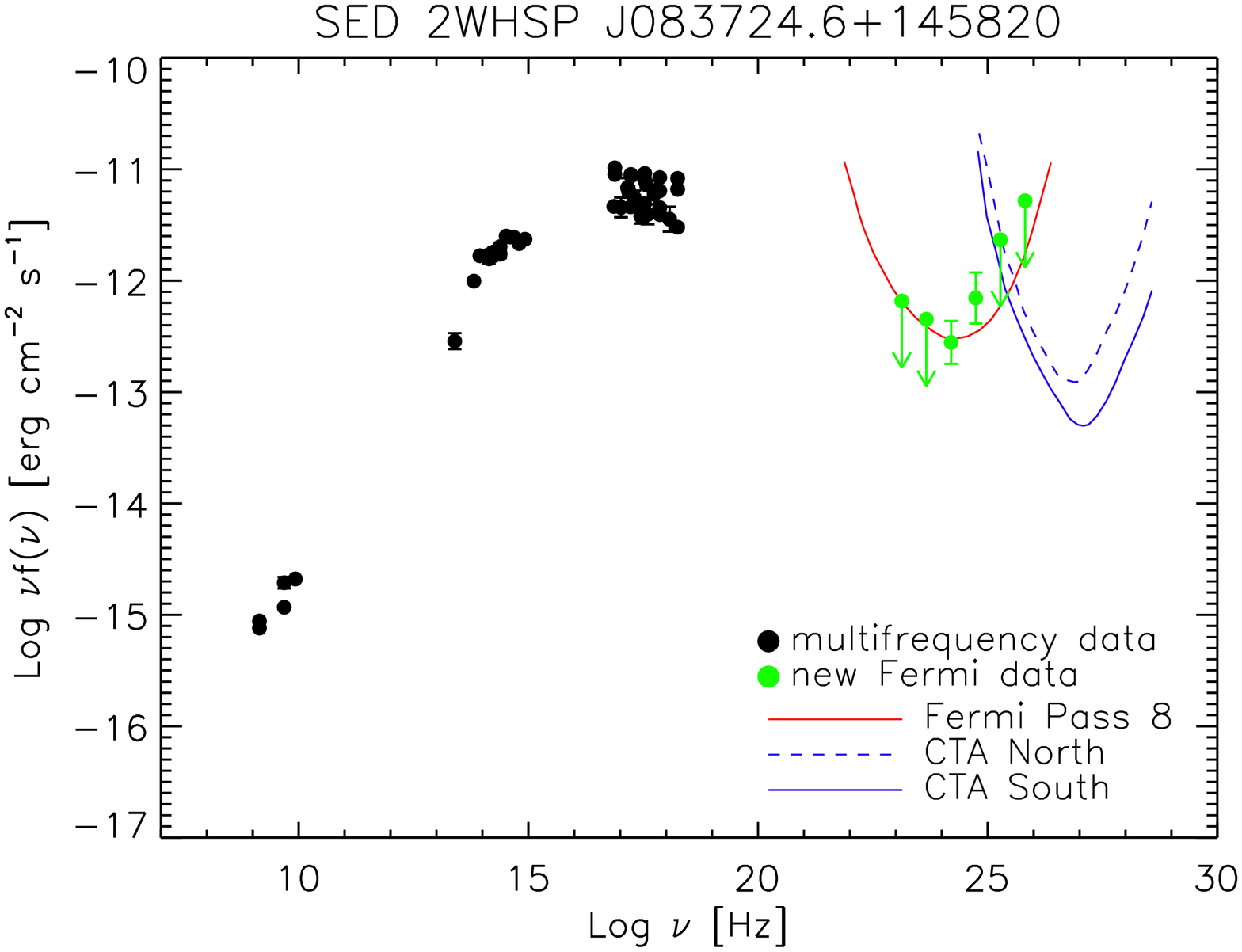} 
\includegraphics[width=0.78\linewidth,angle=90]{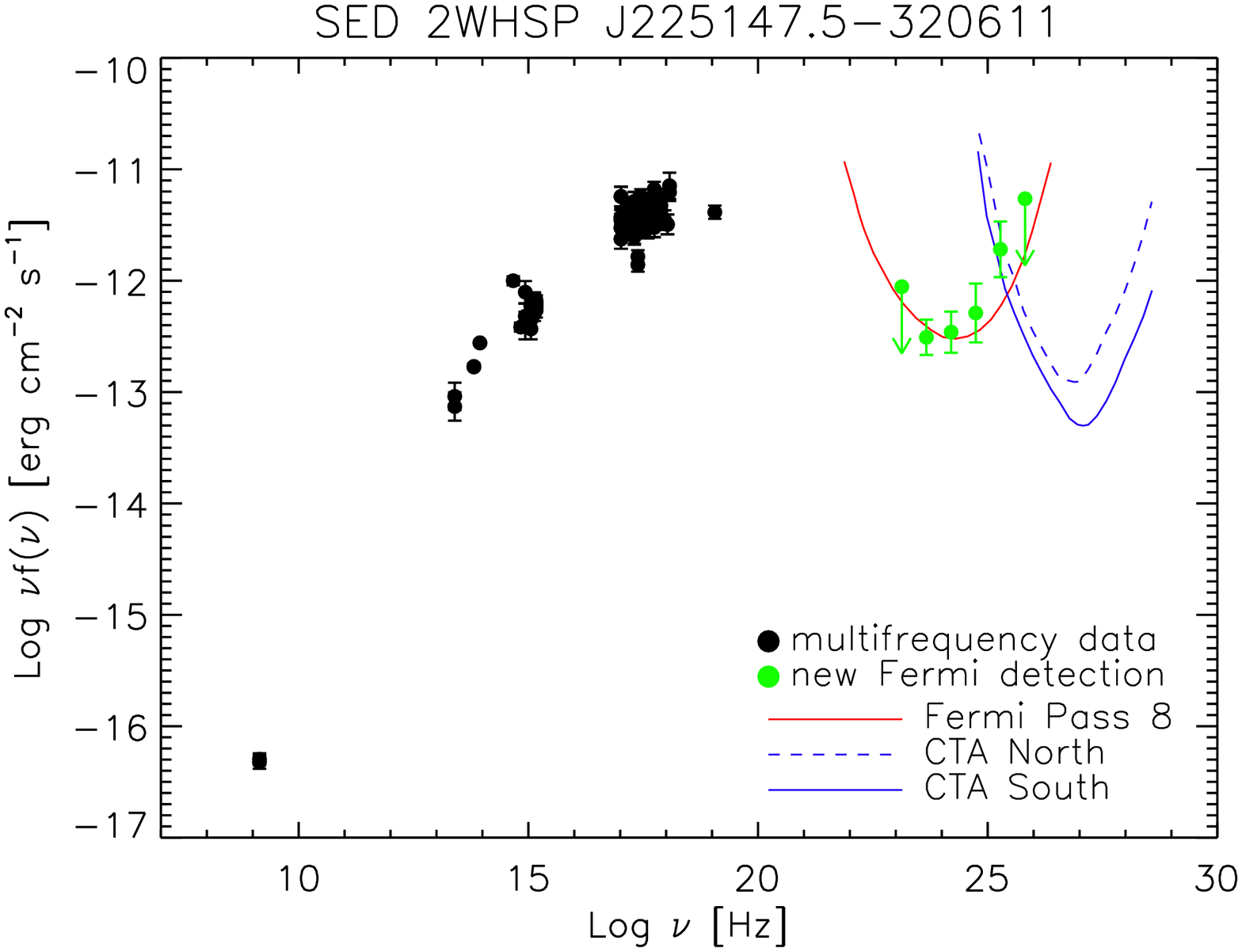}
\caption[]{VHE observations candidates. Top: 2WHSP J083724.6+145820; bottom: 
2WHSP J225147.5$-$320611. The red line and blue lines are the {\it Fermi} Pass 8 and CTA sensitivities, respectively. The green circles are the data from {\it Fermi} Pass 8, and the black points are the data from other wavebands. The Pass 8 data are obtained from the {\it Fermi} tool using the 2WHSP position. These sources are not in the 3FGL catalog yet \citep[see][]{Arsioli2016}.}
\label{gammanew}
\end{figure}

For example, 2WHSP J083724.6+145820 (see Fig.~\ref{gammanew}), has \nupeak~$\sim 10^{16.7}$~Hz and \nufnupeak$\sim 10^{-11}~\rm{erg}~\rm{cm}^{-2}~\rm{s}^{-1}$ (or FOM$=2$), but it had no $\gamma$-ray counterpart until recently. 
The green points in Fig.~\ref{gammanew} correspond to the new $\gamma$-ray data 
presented in \citet{Arsioli2016}. 
Another example is 2WHSP J225147.5$-$320611, which has \nupeak~$ > 10^{18}$~Hz  and 
\nufnupeak~$>10^{-11.3}~\rm{erg}~\rm{cm}^{-2}~\rm{s}^{-1}$ (FOM~$>1$), but also had 
no $\gamma$-ray counterpart in current available $\gamma$-ray or VHE catalogs (1/2/3 FGL, 1/2 FHL, and TeVCat) until it was detected by \cite{Arsioli2016} thanks to the 2WHSP, which points to promising x-ray targets. 

To better assess the percentage of detection of HSP blazars in the \gr\, band, in fact, \cite{Arsioli2016} have recently performed a dedicated $\gamma$-ray analysis of all 2WHSP sources with FOM~$\geq 0.16$, using archival {\it Fermi}-LAT observations integrated over 7.2 years of observations. 
By using the position of 2WHSP sources as seeds for the data analysis, $\approx 85$ sources were identified at the $> 5\sigma~({\rm TS}>25)$ level, and another 65 at a less significant ($10<{\rm TS}<25$) level.  
These results demonstrate the potential of HSP catalogs for the detection and identification of \gr\, and VHE sources. 

Apart from that, the CTA flux limit/sensitivity could be as low as $3\times10^{-13}~\rm{erg}~\rm{cm}^{-2}~\rm{s}^{-1}$ \citep{Rieger2013} or $\sim 1$~mCrab at 1~TeV for 50-hour exposure. 
Clearly, from  Fig.~\ref{gammanew}, 2WHSP J083724.6+145820 and 2WHSP J225147.5$-$320611 may be detected by CTA in the future (since they are above the CTA sensitivity for exposure time 50 hours, the blue lines). 
Therefore, with the benefit of multi-wavelength work, we provide here many candidates for future VHE observations. 

\subsection{HSP blazars as neutrino and cosmic ray emitters?}
\label{neutrinocand} 
Blazars have been considered as likely neutrino sources for
quite some time \citep[e.g.][]{Mannheim1995}. 
\citet{Padovani2014} have suggested that blazars of the HSP type, where particles are accelerated to the highest energies, 
may be good candidates for neutrino emission and presented evidence for an association between HSP blazars and neutrinos detected 
by the IceCube South Pole Neutrino Observatory\footnote{http://icecube.wisc.edu}.

\citet{Petropoulou2015} further modelled the HE SED of six HSPs selected by \citet{Padovani2014} 
as most probable neutrino sources and predicted their neutrino fluxes. All six predicted fluxes were consistent, within the errors, with the observed
neutrino fluxes from IceCube, especially so for two sources (MKN421 and H1914$-$194). 

\citet{Padovani2016} have recently cross-matched two VHE catalogs and the 2WHSP with the most recent IceCube neutrino lists 
\citep{IceCube2015}, measuring the number of neutrino events with at least one $\gamma$-ray counterpart. 
In all three catalogs they observed a positive fluctuation with respect to the mean random expectation at a significance level between 
0.4 and 1.3\%, with a p-value of 0.7\% for 2WHSP sources with FOM~$\ga 1$. 
{\it All} HBLs considered to be the most probable counterparts of IceCube neutrinos are 2WHSP sources, which strongly suggests that 
strong, VHE $\gamma$-ray HBLs are so far the most promising blazar counterparts of astrophysical neutrinos.   

Finally, \cite{Resconi2016} have presented evidence of a direct connection between HSP, very high energy neutrinos, and ultra high energy cosmic rays (UHECRs) by correlating the same catalogs used by \citet{Padovani2016} with UHECRs from the Pierre Auger Observatory and the Telescope Array. 
A maximal excess of 80 cosmic rays (41.9 expected) was observed for 2FHL HBL. 
The chance probability for this to happen is $1.6 \times 10^{-5}$, which translates to $5.5 \times 10^{-4}$ (3.26$\sigma$) after compensation for trials. 
 

\section{Conclusions}
\label{conclusion}

We have  assembled the 2WHSP catalog, currently the largest and most complete existing catalog 
of HSP blazars, by using a multi-frequency method and a detailed comparison with existing lists 
of \gr\, emitting blazars. 2WHSP extends the previous 1WHSP catalog \citep{Arsioli2015a} down to 
lower Galactic latitudes  ($ |b|>10^{\circ} $) and to fainter IR fluxes. In addition, it includes all the 
bright known HSP blazars close to the Galactic plane. The 2WHSP sample includes 1,693 
confirmed or candidates HSP blazars and was also put together to provide a large list of 
potential targets for VHE and multi-messenger observations. 


The average \nupeak\, for our catalog is $ \langle\log\nu_{\rm peak}\rangle = 16.22 \pm 0.02$ Hz and the average redshift 
is $\langle z \rangle=0.331 \pm 0.008$. 
We have shown that the SWCD region needs to be extended to include HSPs in which the host galaxy is dominant. 

Our radio $\log$N-$\log$S shows that the number of HSP blazars over the whole sky is $> 2,000$ and that HBL
make up $\sim 10\%$ of all BL Lacs. 


Finally, we note that this catalog has already been used to provide seeds for the identification 
of new {\it Fermi}-LAT objects and to look for astrophysical counterparts to neutrino and 
UHECR sources \citep{Padovani2016,Resconi2016}, which proves the relevance of having a large
HSP catalog for multi-messenger astronomy.



\begin{acknowledgements}
YLC is supported by the Government of the Republic of China (Taiwan), BA is supported by the 
Brazilian Scientific Program Ci$\hat{\rm{e}}$ncias sem Fronteiras - Cnpq. This work was supported by the 
Agenzia Spaziale Italiana Science Data Center (ASDC) and the University La Sapienza of Rome, 
Department of Physics. PP thanks the ASDC for the hospitality and partial 
financial support for his visits. We made use of archival data and bibliographic information obtained 
from the NASA/IPAC Extragalactic Database (NED), data and software facilities from the ASDC 
managed by the Italian Space Agency (ASI). Extensive use was made of the TOPCAT software package (http://www.star.bris.ac.uk/$\sim$mbt/topcat/). 

\end{acknowledgements}


\bibliographystyle{aa}
\bibliography{2whsp}

\begin{thebibliography}{70}
\expandafter\ifx\csname natexlab\endcsname\relax\def\natexlab#1{#1}\fi

\bibitem[{Abdo {et~al.}(2010)Abdo, Ackermann, Agudo, Ajello, Aller, Aller,
  Angelakis, Arkharov, Axelsson, Bach, \& al.}]{Abdo2010}
Abdo, A., Ackermann, M., Agudo, I., {et~al.} 2010, ApJ, 716, 30

\bibitem[{Ackermann {et~al.}(2016)Ackermann, Ajello, Atwood, Baldini, Ballet,
  Barbiellini, Bastieri, {Becerra Gonzalez}, Bellazzini, Bissaldi, Blandford,
  Bloom, Bonino, Bottacini, Brandt, Bregeon, Bruel, Buehler, Buson, Caliandro,
  Cameron, Caputo, Caragiulo, Caraveo, Cavazzuti, Cecchi, Charles, Chekhtman,
  Cheung, Chiang, Chiaro, Ciprini, Cohen, Cohen-Tanugi, Cominsky, Conrad,
  Cuoco, Cutini, D'Ammando, de~Angelis, de~Palma, Desiante, {Di Mauro}, {Di
  Venere}, Dom{\'{i}}nguez, Drell, Favuzzi, Fegan, Ferrara, Focke, Fortin,
  Franckowiak, Fukazawa, Funk, Furniss, Fusco, Gargano, Gasparrini, Giglietto,
  Giommi, Giordano, Giroletti, Glanzman, Godfrey, Grenier, Grondin, Guillemot,
  Guiriec, Harding, Hays, Hewitt, Hill, Horan, Iafrate, Hartmann, Jogler,
  J{\'{o}}hannesson, Johnson, Kamae, Kataoka, Kn{\"{o}}dlseder, Kuss, {La
  Mura}, Larsson, Latronico, Lemoine-Goumard, Li, Li, Longo, Loparco, Lott,
  Lovellette, Lubrano, Madejski, Maldera, Manfreda, Mayer, Mazziotta,
  Michelson, Mirabal, Mitthumsiri, Mizuno, Moiseev, Monzani, Morselli,
  Moskalenko, Murgia, Nuss, Ohsugi, Omodei, Orienti, Orlando, Ormes, Paneque,
  Perkins, Pesce-Rollins, Petrosian, Piron, Pivato, Porter, Rain{\`{o}}, Rando,
  Razzano, Razzaque, Reimer, Reimer, Reposeur, Romani, S{\'{a}}nchez-Conde,
  {Saz Parkinson}, Schmid, Schulz, Sgr{\`{o}}, Siskind, Spada, Spandre,
  Spinelli, Suson, Tajima, Takahashi, Takahashi, Takahashi, Thayer, Thompson,
  Tibaldo, Torres, Tosti, Troja, Vianello, Wood, Wood, Yassine, Zaharijas, \&
  Zimmer}]{Fermi2fhl2015}
Ackermann, M., Ajello, M., Atwood, W.~B., {et~al.} 2016, ApJS, 222, 5

\bibitem[{Ajello {et~al.}(2015)Ajello, Gasparrini, Sanchez-Conde, Zaharijas,
  Gustafsson, Cohen-Tanugi, Dermer, Inoue, Hartmann, Ackermann, Bechtol,
  Franckowiak, Reimer, Romani, \& Strong}]{Ajello2015}
Ajello, M., Gasparrini, D., Sanchez-Conde, M., {et~al.} 2015, ApJ, 800, L27

\bibitem[{Antonucci(1993)}]{Antonucci1993}
Antonucci, R. 1993, ARA{\&}A, 31, 473

\bibitem[{Arsioli \& Chang(2016)}]{Arsioli2016}
Arsioli, B. \& Chang, Y.-L. 2016, A{\&}A, submitted

\bibitem[{Arsioli {et~al.}(2015)Arsioli, Fraga, Giommi, Padovani, \&
  Marrese}]{Arsioli2015a}
Arsioli, B., Fraga, B., Giommi, P., Padovani, P., \& Marrese, P.~M. 2015,
  A{\&}A, 579, A34

\bibitem[{Blandford \& Rees(1978)}]{Blandford1978}
Blandford, R. \& Rees, M. 1978, in Pittsburgh Conf. BL Lac Objects, ed. A.~M.
  Wolfe (Pittsburgh: University of Pittsburgh press), 341--347

\bibitem[{B{\"{o}}hringer \& Werner(2010)}]{Bohringer2010}
B{\"{o}}hringer, H. \& Werner, N. 2010, A{\&}ARv, 127

\bibitem[{Bonnoli {et~al.}(2015)Bonnoli, Tavecchio, Ghisellini, \&
  Sbarrato}]{Bonnoli2015}
Bonnoli, G., Tavecchio, F., Ghisellini, G., \& Sbarrato, T. 2015, MNRAS, 451,
  611

\bibitem[{Cao {et~al.}(2010)Cao, Bi, Cao, Chan, Chen, He, Chen, Chen, Jiang,
  He, Hu, Huang, Li, Liu, Lu, Ma, Ma, Sheng, Wang, Wang, Wu, Xiao, Yao, Zhang,
  Zhu, An, Li, Zhao, Liu, Sun, Zhao, Huang, Dai, Yang, Zhang, Ma, Mao, Xu,
  Feng, Li, Xue, Zhang, Feng, Jia, Zhou, Zhu, \& {Chen, T. L.
  Danzengluobu;Yuan, A. F.; Cui, S. W.; Zhang}}]{Cao2010}
Cao, Z., Bi, X.~J., Cao, Z., {et~al.} 2010, in 38th COSPAR Sci. Assem., Bremen

\bibitem[{Condon {et~al.}(1998)Condon, Cotton, Greisen, Yin, Perley, Taylor, \&
  Broderick}]{Condon1998}
Condon, J., Cotton, W., Greisen, E., {et~al.} 1998, AJ, 115, 1693

\bibitem[{Costamante {et~al.}(2001)Costamante, Ghisellini, Giommi, Tagliaferri,
  Celotti, Chiaberge, Fossati, Maraschi, Tavecchio, Treves, \&
  Wolter}]{Costamante2001}
Costamante, L., Ghisellini, G., Giommi, P., {et~al.} 2001, A{\&}A, 371, 512

\bibitem[{{Cutri} {et~al.}(2013){Cutri}, {Wright}, {Conrow}, {Fowler},
  {Eisenhardt}, {Grillmair}, {Kirkpatrick}, {Masci}, {McCallon}, {Wheelock},
  {Fajardo-Acosta}, {Yan}, {Benford}, {Harbut}, {Jarrett}, {Lake}, {Leisawitz},
  {Ressler}, {Stanford}, {Tsai}, {Liu}, {Helou}, {Mainzer}, {Gettings},
  {Gonzalez}, {Hoffman}, {Marsh}, {Padgett}, {Skrutskie}, {Beck}, {Papin}, \&
  {Wittman}}]{Cutri2013}
{Cutri}, R.~M., {Wright}, E.~L., {Conrow}, T., {et~al.} 2013, {Explanatory
  Supplement to the AllWISE Data Release Products}, Tech. rep.

\bibitem[{D'Abrusco {et~al.}(2012)D'Abrusco, Massaro, Ajello, Grindlay, Smith,
  \& Tosti}]{DAbrusco2012}
D'Abrusco, R., Massaro, F., Ajello, M., {et~al.} 2012, ApJ, 748, 68

\bibitem[{{Danforth} {et~al.}(2010){Danforth}, {Keeney}, {Stocke}, {Shull}, \&
  {Yao}}]{PG1553}
{Danforth}, C.~W., {Keeney}, B.~A., {Stocke}, J.~T., {Shull}, J.~M., \& {Yao},
  Y. 2010, \apj, 720, 976

\bibitem[{D'Elia {et~al.}(2013)D'Elia, Perri, Puccetti, Capalbi, Giommi,
  Burrows, Campana, Tagliaferri, Cusumano, Evans, Gehrels, Kennea, Moretti,
  Nousek, Osborne, Romano, \& Stratta}]{DElia2013}
D'Elia, V., Perri, M., Puccetti, S., {et~al.} 2013, A{\&}A, 551, A142

\bibitem[{Dermer {et~al.}(2011)Dermer, Cavadini, Razzaque, Finke, Chiang, \&
  Lott}]{Dermer2011}
Dermer, C.~D., Cavadini, M., Razzaque, S., {et~al.} 2011, ApJ, 733, L21

\bibitem[{{Di Mauro} {et~al.}(2014){Di Mauro}, Donato, Lamanna, Sanchez, \&
  Serpico}]{DiMauro2014}
{Di Mauro}, M., Donato, F., Lamanna, G., Sanchez, D.~A., \& Serpico, P.~D.
  2014, ApJ, 786, 129

\bibitem[{Elvis {et~al.}(1992)Elvis, Plummer, Schachter, \&
  Fabbiano}]{Elvis1992}
Elvis, M., Plummer, D., Schachter, J., \& Fabbiano, G. 1992, ApJS, 80, 257

\bibitem[{Evans {et~al.}(2010)Evans, Primini, Glotfelty, Anderson, Bonaventura,
  Chen, Davis, Doe, Evans, Fabbiano, Galle, Gibbs, Grier, Hain, Hall, Harbo,
  He, Houck, Karovska, Kashyap, Lauer, McCollough, McDowell, Miller, Mitschang,
  Morgan, Mossman, Nichols, Nowak, Plummer, Refsdal, Rots, Siemiginowska,
  Sundheim, Tibbetts, {Van Stone}, Winkelman, \& Zografou}]{Evans2010}
Evans, I.~N., Primini, F.~A., Glotfelty, K.~J., {et~al.} 2010, ApJS, 189, 37

\bibitem[{Finke {et~al.}(2015)Finke, {Reyes, Luis C. Georganopoulos, Markos
  Reynolds}, Ajello, Fegan, \& McCann}]{Finke2015}
Finke, J.~D., {Reyes, Luis C. Georganopoulos, Markos Reynolds}, K., Ajello, M.,
  Fegan, S.~J., \& McCann, K. 2015, ApJ, 814, 20

\bibitem[{{Furniss} {et~al.}(2013){Furniss}, {Williams}, {Danforth},
  {Fumagalli}, {Prochaska}, {Primack}, {Urry}, {Stocke}, {Filippenko}, \&
  {Neely}}]{pks1424p240HighZ}
{Furniss}, A., {Williams}, D.~A., {Danforth}, C., {et~al.} 2013, \apjl, 768,
  L31

\bibitem[{{Giommi} {et~al.}(2002){Giommi}, {Capalbi}, {Fiocchi}, {Memola},
  {Perri}, {Piranomonte}, {Rebecchi}, \& {Massaro}}]{Giommi2002}
{Giommi}, P., {Capalbi}, M., {Fiocchi}, M., {et~al.} 2002, in Blazar
  Astrophysics with BeppoSAX and Other Observatories, ed. P.~{Giommi},
  E.~{Massaro}, \& G.~{Palumbo}, 63

\bibitem[{Giommi {et~al.}(2006)Giommi, Colafrancesco, Cavazzuti, Perri, \&
  Pittori}]{Giommi2006}
Giommi, P., Colafrancesco, S., Cavazzuti, E., Perri, M., \& Pittori, C. 2006,
  A{\&}A, 445, 843

\bibitem[{Giommi {et~al.}(2009)Giommi, Colafrancesco, Padovani, Gasparrini,
  Cavazzuti, \& Cutini}]{Giommi2009}
Giommi, P., Colafrancesco, S., Padovani, P., {et~al.} 2009, A{\&}A, 508, 107

\bibitem[{Giommi {et~al.}(1999)Giommi, Menna, \& Padovani}]{Giommi1999}
Giommi, P., Menna, M.~T., \& Padovani, P. 1999, MNRAS, 310, 465

\bibitem[{Giommi \& Padovani(2015)}]{Giommi2015}
Giommi, P. \& Padovani, P. 2015, MNRAS, 450, 2404

\bibitem[{Giommi {et~al.}(2012)Giommi, Padovani, Polenta, Turriziani, D'Elia,
  \& Piranomonte}]{Giommi2012a}
Giommi, P., Padovani, P., Polenta, G., {et~al.} 2012, MNRAS, 420, 2899

\bibitem[{Giommi {et~al.}(2005)Giommi, Piranomonte, Perri, \&
  Padovani}]{Giommi2005}
Giommi, P., Piranomonte, S., Perri, M., \& Padovani, P. 2005, A{\&}A, 434, 385

\bibitem[{Harris {et~al.}(1993)Harris, Forman, Gioia, Hale, {Harnden, F. R.},
  Jones, Karakashian, Maccacaro, McSweeney, \& Primini}]{Harris1993}
Harris, D.~E., Forman, W., Gioia, I.~M., {et~al.} 1993, {The Einstein
  Observatory catalog of IPC X ray sources. Volume 1E: Documentation}

\bibitem[{{IceCube Collaboration}(2015)}]{IceCube2015}
{IceCube Collaboration}. 2015, in 34th Int. Cosm. Ray Conf.

\bibitem[{Kapanadze(2013)}]{Kapanadze2013}
Kapanadze, B. 2013, AJ, 145, 31

\bibitem[{Manch {et~al.}(2003)Manch, Murphy, Buttery, Curran, Hunstead,
  Piestrzynski, Robertson, \& Sadler}]{Manch2003}
Manch, T., Murphy, T., Buttery, H.~J., {et~al.} 2003, MNRAS, 342, 1117

\bibitem[{Mannheim(1995)}]{Mannheim1995}
Mannheim, K. 1995, APh, 3, 295

\bibitem[{{Masetti} {et~al.}(2013){Masetti}, {Sbarufatti}, {Parisi},
  {Jim{\'e}nez-Bail{\'o}n}, {Chavushyan}, {Vogt}, {Sguera}, {Stephen},
  {Palazzi}, {Bassani}, {Bazzano}, {Fiocchi}, {Galaz}, {Landi}, {Malizia},
  {Minniti}, {Morelli}, \& {Ubertini}}]{masetti2013}
{Masetti}, N., {Sbarufatti}, B., {Parisi}, P., {et~al.} 2013, \aap, 559, A58

\bibitem[{Massaro {et~al.}(2009)Massaro, Giommi, Leto, Marchegiani, Maselli,
  Perri, Piranomonte, \& Sclavi}]{Massaro2009}
Massaro, E., Giommi, P., Leto, C., {et~al.} 2009, A{\&}A, 495, 691

\bibitem[{Massaro {et~al.}(2015)Massaro, Maselli, Leto, Marchegiani, Perri,
  Giommi, \& Piranomonte}]{Massaro2015}
Massaro, E., Maselli, A., Leto, C., {et~al.} 2015, Ap{\&}SS, 357, 75

\bibitem[{Massaro {et~al.}(2012)Massaro, Nesci, \& Piranomonte}]{Massaro2012}
Massaro, E., Nesci, R., \& Piranomonte, S. 2012, MNRAS, 422, 2322

\bibitem[{Massaro {et~al.}(2011)Massaro, D'Abrusco, Ajello, Grindlay, \&
  Smith}]{Massaro2011}
Massaro, F., D'Abrusco, R., Ajello, M., Grindlay, J., \& Smith, H. 2011, ApJ,
  740, L48

\bibitem[{Padovani {et~al.}(1993)Padovani, Ghisellini, Fabian, \&
  Celotti}]{Padovani1993}
Padovani, P., Ghisellini, G., Fabian, A.~C., \& Celotti, A. 1993, MNRAS, 260,
  L21

\bibitem[{Padovani \& Giommi(1995)}]{Padovani1995}
Padovani, P. \& Giommi, P. 1995, ApJ, 444, 567

\bibitem[{Padovani \& Giommi(2015)}]{Padovani2015a}
Padovani, P. \& Giommi, P. 2015, MNRAS, 446, L41

\bibitem[{Padovani {et~al.}(2007)Padovani, Giommi, Landt, \&
  Perlman}]{Padovani2007}
Padovani, P., Giommi, P., Landt, H., \& Perlman, E.~S. 2007, ApJ, 662, 182

\bibitem[{Padovani \& Resconi(2014)}]{Padovani2014}
Padovani, P. \& Resconi, E. 2014, MNRAS, 443, 474

\bibitem[{Padovani {et~al.}(2016)Padovani, Resconi, Giommi, Arsioli, \&
  Chang}]{Padovani2016}
Padovani, P., Resconi, E., Giommi, P., Arsioli, B., \& Chang, Y.~L. 2016,
  MNRAS, 457, 3582

\bibitem[{Panzera {et~al.}(2003)Panzera, Campana, Covino, Lazzati, Mignani,
  Moretti, \& Tagliaferri}]{Panzera2003}
Panzera, M.~R., Campana, S., Covino, S., {et~al.} 2003, Astron. Astrophys.,
  399, 351

\bibitem[{P{\'{e}}rez-Torres {et~al.}(2009)P{\'{e}}rez-Torres, Zandanel,
  Guerrero, Pal, Profumo, Prada, \& Panessa}]{PerezTorres2009}
P{\'{e}}rez-Torres, M.~A., Zandanel, F., Guerrero, M.~A., {et~al.} 2009, MNRAS,
  396, 2237

\bibitem[{Petropoulou {et~al.}(2015)Petropoulou, Dimitrakoudis, Padovani,
  Mastichiadis, \& Resconi}]{Petropoulou2015}
Petropoulou, M., Dimitrakoudis, S., Padovani, P., Mastichiadis, A., \& Resconi,
  E. 2015, MNRAS, 448, 2412

\bibitem[{Pfrommer {et~al.}(2013)Pfrommer, Broderick, Chang, Puchwein, \&
  Springel}]{Pfrommer2013}
Pfrommer, C., Broderick, A.~E., Chang, P., Puchwein, E., \& Springel, V. 2013,
  ArXiv e-prints

\bibitem[{{Pian} {et~al.}(1998){Pian}, {Vacanti}, {Tagliaferri}, {Ghisellini},
  {Maraschi}, {Treves}, {Urry}, {Fiore}, {Giommi}, {Palazzi}, {Chiappetti}, \&
  {Sambruna}}]{Pian1998}
{Pian}, E., {Vacanti}, G., {Tagliaferri}, G., {et~al.} 1998, \apjl, 492, L17

\bibitem[{Piranomonte {et~al.}(2007)Piranomonte, Perri, Giommi, Landt, \&
  Padovani}]{Piranomonte2007}
Piranomonte, S., Perri, M., Giommi, P., Landt, H., \& Padovani, P. 2007,
  A{\&}A, 470, 787

\bibitem[{{Pita} {et~al.}(2014){Pita}, {Goldoni}, {Boisson}, {Lenain}, {Punch},
  {G{\'e}rard}, {Hammer}, {Kaper}, \& {Sol}}]{pita2013}
{Pita}, S., {Goldoni}, P., {Boisson}, C., {et~al.} 2014, \aap, 565, A12

\bibitem[{{Planck Collaboration} {et~al.}(2015){Planck Collaboration}, Ade,
  Aghanim, Arnaud, \& Ashdown}]{Planck2015}
{Planck Collaboration}, Ade, P., Aghanim, N., Arnaud, M., \& Ashdown, M. 2015,
  Astron. Astrophys., 581, A14

\bibitem[{Puccetti {et~al.}(2011)Puccetti, Capalbi, Giommi, Perri, Stratta,
  Angelini, Burrows, Campana, Chincarini, Cusumano, Gehrels, Moretti, Nousek,
  Osborne, \& Tagliaferri}]{Puccetti2011}
Puccetti, S., Capalbi, M., Giommi, P., {et~al.} 2011, A{\&}A, 528, A122

\bibitem[{Resconi {et~al.}(2016)Resconi, Coenders, Padovani, Giommi, \&
  Caccianiga}]{Resconi2016}
Resconi, E., Coenders, S., Padovani, P., Giommi, P., \& Caccianiga, L. 2016,
  Phys. Review Lett., submitted

\bibitem[{Rieger {et~al.}(2013)Rieger, de~O{\~{n}}a-Wilhelmi, \&
  Aharonian}]{Rieger2013}
Rieger, F.~M., de~O{\~{n}}a-Wilhelmi, E., \& Aharonian, F.~A. 2013, FrPhy, 8,
  714

\bibitem[{Rosen {et~al.}(2016)Rosen, Webb, Watson, Ballet, Barret, Braito,
  Carrera, Ceballos, Coriat, {Della Ceca}, Denkinson, Esquej, Farrell,
  Freyberg, Gris{\'{e}}, Guillout, Heil, Law-Green, Lamer, Lin, Martino,
  Michel, Motch, Gomez-Moran, Page, Page, Page, Pakull, Pye, Read, Rodriguez,
  Sakano, Saxton, Schwope, Scott, Sturm, Traulsen, Yershov, \&
  Zolotukhin}]{Rosen2015}
Rosen, S.~R., Webb, N.~A., Watson, M.~G., {et~al.} 2016, A{\&}A, 590, A1

\bibitem[{Rybicki \& Lightman(1986)}]{Rybicki1986}
Rybicki, G.~B. \& Lightman, A.~P. 1986, {Radiative Processes in Astrophysics}
  (Wiley-VCH)

\bibitem[{Sarazin(1988)}]{Sarazin1988}
Sarazin, C.~L. 1988, {X-ray emission from clusters of galaxies} (Cambridge:
  Cambridge Astrophysics Series)

\bibitem[{Saxton {et~al.}(2008)Saxton, Read, Esquej, Freyberg, Altieri, \&
  Bermejo}]{Saxton2008}
Saxton, R.~D., Read, a.~M., Esquej, P., {et~al.} 2008, A{\&}A, 480, 611

\bibitem[{{Sbarufatti} {et~al.}(2005){Sbarufatti}, {Treves}, {Falomo}, {Heidt},
  {Kotilainen}, \& {Scarpa}}]{bll}
{Sbarufatti}, B., {Treves}, A., {Falomo}, R., {et~al.} 2005, \aj, 129, 559

\bibitem[{Schachter {et~al.}(1993)Schachter, Stocke, Perlman, Elvis, Remillard,
  Granados, Luu, Huchra, Humphreys, Urry, \& Wallin}]{Schachter1993}
Schachter, J.~F., Stocke, J.~T., Perlman, E., {et~al.} 1993, ApJ, 412, 541

\bibitem[{{Shaw} {et~al.}(2013{\natexlab{a}}){Shaw}, {Filippenko}, {Romani},
  {Cenko}, \& {Li}}]{shaw2}
{Shaw}, M.~S., {Filippenko}, A.~V., {Romani}, R.~W., {Cenko}, S.~B., \& {Li},
  W. 2013{\natexlab{a}}, \aj, 146, 127

\bibitem[{{Shaw} {et~al.}(2013{\natexlab{b}}){Shaw}, {Romani}, {Cotter},
  {Healey}, {Michelson}, {Readhead}, {Richards}, {Max-Moerbeck}, {King}, \&
  {Potter}}]{bllacz2}
{Shaw}, M.~S., {Romani}, R.~W., {Cotter}, G., {et~al.} 2013{\natexlab{b}},
  \apj, 764, 135

\bibitem[{Stocke {et~al.}(1991)Stocke, Morris, Gioia, Maccacaro, Schild,
  Wolter, Fleming, \& Henry}]{Stocke1991}
Stocke, J.~T., Morris, S.~L., Gioia, I.~M., {et~al.} 1991, ApJS, 76, 813

\bibitem[{Urry \& Padovani(1995)}]{Urry1995}
Urry, C.~M. \& Padovani, P. 1995, PASP, 107, 803

\bibitem[{Voges {et~al.}(1999)Voges, Aschenbach, Boller, Br{\"{a}}uninger,
  Briel, Burkert, Dennerl, Englhauser, Gruber, Haberl, Hartner, Hasinger,
  K{\"{u}}rster, Pfeffermann, Pietsch, Predehl, Rosso, Schmitt, Tr{\"{u}}mper,
  \& Zimmermann}]{Voges1999}
Voges, W., Aschenbach, B., Boller, T., {et~al.} 1999, A{\&}A, 349, 389

\bibitem[{Voges {et~al.}(2000)Voges, Aschenbach, Boller, Brauninger, Briel,
  Burkert, Dennerl, Englhauser, Gruber, Haberl, Hartner, Hasinger, Pfeffermann,
  Pietsch, Predehl, Schmitt, Trumper, \& Zimmermann}]{Voges2000}
Voges, W., Aschenbach, B., Boller, T., {et~al.} 2000, in IAU Circ.

\bibitem[{White {et~al.}(2000)White, Giommi, \& Angelini}]{White2000}
White, N., Giommi, P., \& Angelini, L. 2000, VizieR Online Data Cat.

\bibitem[{White {et~al.}(1997)White, Becker, Helfand, \& Gregg}]{White1997}
White, R.~L., Becker, R.~H., Helfand, D.~J., \& Gregg, M.~D. 1997, ApJ, 475,
  479

\end{thebibliography}

\begin{center}
\begin{longtable}
{>{\raggedright\arraybackslash}p{2.6cm}>{\centering\arraybackslash}p{0.9cm}>{\centering\arraybackslash}p{1.4cm}>{\centering\arraybackslash}p{2cm}>{\centering\arraybackslash}p{0.8cm}>{\centering\arraybackslash}p{3.6cm}>{\centering\arraybackslash}p{1.4cm}>{\centering\arraybackslash}p{1.7cm}>{\centering\arraybackslash}p{0.5cm}}
\caption{\label{table1} 2WHSP sample. The first column shows the source name in the 2WHSP catalog, which are based on the IR position (hms-J2000) from the WISE catalog. The column 5BZCat gives the source class: BZB-BZQ-BZU, and sources not in this catalog are considered new blazars (if optical identification is available) or blazar-cadidates (if  optical identification is not available). Upper/lower limits on  redshift are taken from \cite{bllacz2} (a), \cite{pita2013} (b), \cite{pks1424p240HighZ} (c), \cite{PG1553} (d), \cite{shaw2} (e), \cite{masetti2013} (f), \cite{bll} (g), or calculated by us (as described in \cite{Arsioli2015a}). The column {\it Fermi}-LAT shows the 2WHSP $\gamma$-ray counterpart in the {\it Fermi} catalogs 1/2/3FGL (0.1-100GeV), and also lists complementary $\gamma$-ray detections (0.3-500GeV) from \cite{Arsioli2016}; their respective photon indexes and uncertainties are listed in column $\Gamma$. The column 2FHL list sources detected at $>$50GeV from \cite{Fermi2fhl2015}, and the column FOM represents the likelihood of detection in the TeV band (see sec \ref{vhecand}). The lower limits and uncertain values on $\log~\nu_{peak}$ and $\log~\nu f_{\nu}$ are due to limited data, specially in the X-ray band. Entries marked with an asterisk are uncertain. The full catalog is available at www.asdc.asi.it/2whsp containing complementary information and a direct link to the SED building tool where the multi-frequency spectra for each source can be quickly visualized.} \\
\hline
2WHSP J & 5BZCat & Log(\nupeak) & Log(\nufnupeak) & z & {\it Fermi}-LAT & $\Gamma$ & 2FHL J & FOM \\
\hline
\endfirsthead
\caption{continued.}\\
\hline\hline
2WHSP J & 5BZ$_{cat}$ & Log(\nupeak) & Log(\nufnupeak) & z & {\it Fermi}-LAT & $\Gamma$ & 2FHL J & FOM \\
\hline
\endhead
\hline
\endfoot
  000116.3+293534 &   -      &     16.5$^{*}$    &     -12.9$^{*}$    &    -                &     -                     &    -                  &     -                &  0.03  \\            
  000116.9$-$315043 &   -      &     15.9          &     -12.3          &    -                &     -                     &    -                  &     -                &  0.10  \\            
  000132.6$-$415524 &   -      &     15.8          &     -11.7          &    -                &     3FGL J0002.2$-$4152     &    2.09$\pm$0.18      &     -                &  0.40  \\            
  000158.0$-$115047 &   -      &     15.5          &     -12.4          &    -                &     -                     &    -                  &     -                &  0.08  \\            
  000215.1$-$672653 &   -      &     15.9          &     -11.9          &    -                &     3FGL J0002.0-6722     &    1.95$\pm$0.15      &     -                &  0.25  \\            
  000236.0$-$081532 &   -      &     16.8          &     -12.3          &    -                &     -                     &    -                  &     -                &  0.10  \\            
  000319.5$-$524726 &   -      &     17.1          &     -11.9          &    -                &     3FGL J0003.2$-$5246     &    1.90$\pm$0.17      &     -                &  0.25  \\            
  000513.6$-$261438 &   -      &     15.4          &     -12.6          &    -                &     -                     &    -                  &     -                &  0.05  \\
\end{longtable}
\end{center}
\label{2whsptable}

\end{document}